\documentclass[useAMS,usenatbib,usegraphicx]{mn2e} 
\usepackage{amssymb} 
\usepackage{times} 
\usepackage{aas_macros} 
\usepackage[usenames]{color} 
\usepackage{color}
\usepackage{hyperref}

\title[A sectoral dipole mode in TIC~63328020]{A tidally tilted sectoral dipole pulsation mode in the eclipsing binary TIC~63328020 }

\author[S. Rappaport et al.]{S. A. Rappaport$^1$\thanks{E-mail: sar@mit.edu}, D. W. Kurtz$^{2,3}$,  G. Handler$^{4}$, D. Jones$^{5,6}$, L. A. Nelson$^{7}$, H. Saio$^{8}$,  \newauthor J. Fuller$^{9}$, D. L. Holdsworth$^{3}$, A. Vanderburg$^{10}$, J. \v{Z}\'ak$^{11,12,5}$, M. Skarka$^{11,13}$, J. Aiken$^{7}$, \newauthor P.\,F.\,L. Maxted$^{14}$, D. J. Stevens$^{15}$, D. L. Feliz$^{16,17}$, and F. Kahraman Ali\c{c}avu\c{s}$^{18,19}$
\\
$^{1}$Department of Physics, and Kavli Institute for Astrophysics and Space Research, M.I.T., Cambridge, MA 02139, USA\\
$^{2}$Centre for Space Research, Physics Department, North West University, Mahikeng 2745, South Africa\\
$^{3}$Jeremiah Horrocks Institute, University of Central Lancashire, Preston PR1 2HE, UK\\
$^{4}$Nicolaus Copernicus Astronomical Center, Polish Academy of Sciences, ul. Bartycka 18, 00-716, Warszawa, Poland\\
$^5$Instituto de Astrof\'isica de Canarias, E-38205 La Laguna, Tenerife, Spain \\
$^6$Departamento de Astrof\'isica, Universidad de La Laguna, E-38206 La Laguna, Tenerife, Spain\\
$^{7}$Department of Physics and Astronomy, Bishop's University, 2600 College St., Sherbrooke, QC J1M 1Z7 \\
$^8$Astronomical Institute, Graduate School of Science, Tohoku University, Sendai 980-8578, Japan\\
$^{9}$Division of Physics, Mathematics and Astronomy, California Institute of Technology, Pasadena, CA 91125, USA\\
$^{10}$Department of Astronomy, The University of Texas at Austin, 2515 Speedway, Stop C1400, Austin, TX 78712, USA\\
$^{11}$Department of Theoretical Physics and Astrophysics, Masaryk Univesity, Kotl\'a\v{r}sk\'a 2, 60200 Brno, Czech Republic\\
$^{12}$ ESO, Karl-Schwarzschild-str. 2, D-85748 Garching, Germany \\
$^{13}$Astronomical Institute, Czech Academy of Sciences, Fri\v{c}ova 298, 25165, Ond\v{r}ejov, Czech Republic\\
$^{14}$Astrophysics Group, Keele University, Staffordshire, ST5 5BG, UK\\  
$^{15}$Department of Astronomy \& Astrophysics and Center for Exoplanets and Habitable Worlds, PSU, 525 Davey Lab, University Park, PA 16802, USA \\
$^{16}$Department of Physics and Astronomy, Vanderbilt University, Nashville, TN 37235, USA \\  
$^{17}$Department of Physics, Fisk University, 1000 17th Avenue North, Nashville, TN 37208, USA \\
$^{18}$\c{C}anakkale Onsekiz Mart University, Faculty of Sciences and Arts, Physics Department, 17100, \c{C}anakkale, Turkey\\ 
$^{19}$\c{C}anakkale Onsekiz Mart University, Astrophysics Research Center and Ulupõnar Observatory, TR-17100, \c{C}anakkale, Turkey\\ 
}
\date{Accepted XXX. Received YYY; in original form ZZZ}

\pubyear{2020}

\begin{document}
\label{firstpage}
\maketitle

\begin{abstract}
  We report the discovery of the third tidally tilted pulsator, TIC~63328020. Observations with the {\em TESS} satellite reveal binary eclipses with an orbital period of 1.1057 d, and $\delta$ Scuti-type pulsations with a mode frequency of 21.09533\,d$^{-1}$.  This pulsation exhibits a septuplet of orbital sidelobes as well as a harmonic quintuplet. Using the oblique pulsator model, the primary oscillation is identified as a sectoral dipole mode with $l = 1, |m| = 1$. We find the pulsating star to have $M_1 \simeq 2.5 \, {\rm M}_\odot$, $R_1 \simeq 3 \, {\rm R}_\odot$, and $T_{\rm eff,1} \simeq 8000$ K, while the secondary has $M_2 \simeq 1.1 \, {\rm M}_\odot$, $R_2 \simeq 2 \, {\rm R}_\odot$, and $T_{\rm eff,2} \simeq 5600$ K.  Both stars appear to be close to filling their respective Roche lobes. The properties of this binary as well as the tidally tilted pulsations differ from the previous two tidally tilted pulsators, HD74423 and CO Cam, in important ways.  We also study the prior history of this system with binary evolution models and conclude that extensive mass transfer has occurred from the current secondary to the primary.   

\end{abstract} 


\section{Introduction}
\label{sec:intro}

The single-sided or tidally-tilted pulsators are a newly recognized type of pulsating star in close binary systems. In these stars the tidal distortion caused by the companion aligns the pulsation axis of the oscillating star with the tidal axis. This has two important consequences. First, the pulsation axis corotates with the orbit and therefore the stellar pulsation modes are seen at varying aspect, leading to amplitude and phase variations with the orbital phase. Second, the tidal distortion of the pulsating star causes an intrinsically uneven distribution of pulsation amplitude over the stellar surface.

These two facts can be used to our astrophysical advantage. Viewing the stellar pulsation over the full orbital cycle allows us to constrain the orbital inclination, $i$, and the obliquity of the pulsation axis, $\beta$, to the orbital axis.  
The required mathematical framework, the oblique pulsator model, has been developed over the past four decades, starting with \cite{1982MNRAS.200..807K}. Even though it was developed for the rapidly oscillating Ap (roAp) stars, whose pulsation axes are tilted with respect to their pulsation axes due to the stellar magnetic fields, it is readily applicable to tidally tilted pulsators. Since the pulsation axes of those stars are located in the orbital plane, the analyses of the binary-induced variability and of the tidally tilted pulsation mutually constrain each other. The predominant shape of the distorted oscillations, and thus the ``pulsational quantum numbers'' -- the spherical degree $l$ and the azimuthal order $m$ -- can be determined from the variation of the pulsation amplitude and phase over the orbit. In reality, the tidal distortion induces coupling between modes of different spherical degree $l$, modifying the perturbed flux distribution at the surface, and sometimes ``tidally trapping" the oscillations on one side of the star.  The theoretical groundwork for this type of analysis was laid by \citet{2020MNRAS.tmp.2716F}.

Two single-sided pulsators have so far been reported in the literature; both are ellipsoidal variables and each contains at least one $\delta$~Scuti pulsator. Ellipsoidal variables are close binary stars with tidally distorted components. They exhibit light variations as the projected stellar surface area and surface gravity vary towards the direction to a distant observer \citep[e.g.,][]{1985ApJ...295..143M} over the orbital cycle. The $\delta$~Scuti stars, on the other hand, are a common group of short-period pulsators located at the intersection of the classical instability strip with the Main Sequence \citep{1979PASP...91....5B,breger00}. Naturally, some $\delta$~Scuti stars are also located in binary systems \citep[e.g.,][]{2017MNRAS.465.1181L}, with a wide range of phenomenology occurring, for instance pulsators in eclipsing binaries \citep{2017MNRAS.470..915K} or the so-called `heartbeat' stars with tidally excited stellar oscillations in binary systems with eccentric orbits \citep[e.g.,][]{2011ApJS..197....4W}. However, in none of these systems was evidence for tidal effects on the pulsation axes, such as those occurring in the single-sided pulsators, reported.

The first such discovery was HD~74423 \citep{2020NatAs.tmp...45H}, which contains two chemically peculiar stars of the $\lambda$ Bootis type in a 1.58-d orbit that are close to filling their Roche lobes. Although the two components are almost identical, only one of them shows $\delta$ Scuti pulsation. Rather unusual for this type of pulsating star, there is only a single mode of oscillation, and it is not clear which of the two components is the pulsator. Shortly afterwards, \cite{2020MNRAS.494.5118K} reported the discovery of a second such system, CO Cam. The properties of this binary are different from those of HD 74423. Its orbital period is somewhat shorter (1.27 d), the secondary component is spectroscopically undetected, hence considerably less luminous than the pulsating primary, which is far from filling its Roche lobe. The pulsating star in the CO Cam system is also chemically peculiar, but it is a marginal metallic-lined A-F star (often denoted with the spectral classification ``Am:''), and it pulsates in at least four tidally distorted modes.

Both HD~74423 and CO~Cam were designated as ``single-sided pulsators'' because they show enhanced pulsation amplitude on the L$_1$ side of the star facing the secondary, as explained by \citet{2020MNRAS.tmp.2716F}.. In the present paper, we report the discovery of the third single-sided pulsator, TIC~63328020, which is different from the two systems studied earlier.  With the discovery of a sectoral pulsation mode in TIC~63328020, we now refer to these stars generally as `tidally tilted pulsators', where the more specific name, `single-sided pulsators', can still be used for those stars that have strongly enhanced pulsation amplitude on one side of the star. 

In Section \ref{sec:pulsations} we discuss how this object was first noticed in Transiting Exoplanet Survey Satellite ({\em TESS}) data and we present a detailed analysis of the pulsations.  In particular, we show that the pulsations are strongly modulated in amplitude and phase around the orbit, with the peak pulsation amplitudes coinciding in time with the maxima of the ellipsoidal light variations (`ELVs'), hence in quadrature with the eclipses.  In Section~\ref{sec:archive} we present the results of a study of archival data for the spectral energy distribution (`SED') as well as the long-term eclipse timing variations (`ETVs') for the system.  Our radial velocity (RV) data for the system are presented and analysed in Section~\ref{sec:RVs}, while  Section~\ref{sec:mcmc} utilizes the RV and SED data to analyse the system properties.  As a complementary analysis of the system parameters, in Section~\ref{sec:phoebe} we combine the RV data and the {\em TESS} light curve via the {\tt phoebe2} code to derive the system parameters.  Finally, in Section~\ref{sec:mdot} we use a series of {\tt MESA} binary evolution grids to understand the formation and evolutionary history of TIC~63328020.  We conclude that there was almost certainly a prior history of mass transfer in the system and that the roles of the primary and secondary stars have reversed.

\section{Eclipses and Pulsations in TIC~63328020}
\label{sec:pulsations}

TIC~63328020 = NSVS~5856840 was reported as an eclipsing binary by \cite{2008AJ....136.1067H}, the only literature reference to this star. It has an  apparent visual magnitude $ \simeq 12$, and no spectral type was given.  The archival properties of TIC~6338020 are summarised in Section~\ref{magnitudes} below. 

In addition to the eclipses, $\delta$~Scuti-type pulsations were discovered by one of us (DWK) during a visual inspection of  {\em TESS} Sector 15 light curves. During its main two-year mission, $2018 - 2020$, {\em TESS} observed almost the whole sky in a search for transiting extrasolar planets around bright stars ($4< I_c<13$) in a wide red-bandpass filter. The measurements were taken in partly overlapping $24\times96^\circ$ sectors around the ecliptic poles that were observed for two 13.6-d satellite orbital periods each \citep{2015JATIS...1a4003R}.

TIC~63328020 was observed by {\em TESS} in Sectors 15 and 16 in 2-min cadence.  We used the pre-search data conditioned simple aperture photometry (PDCSAP) data downloaded from the Mikulski Archive for Space Telescopes (MAST)\footnote{http://archive.stsci.edu/tess/all\_products.html}. The data have a time span of 51.95\,d with a centre point in time of $t_0 = {\rm BJD}~2458737.34462$\footnote{This $t_0$ was used to begin the analysis, but later changed to the time of pulsation maximum to test the oblique pulsator model. For the assessment of phase errors with nonlinear least-squares fitting, it is important that the $t_0$  chosen is near to the centre of the data set. Since frequency and phase are degenerately coupled in the fitting of sinusoids, when $t_0$ is not the centre of the data set, small changes in frequency result in very large changes in phase, since phase is referenced from $t_0$.}, and comprise 34657 data points after some outliers were removed with inspection by eye. 

\begin{figure*}
\centering
\includegraphics[width=0.9\linewidth,angle=0]{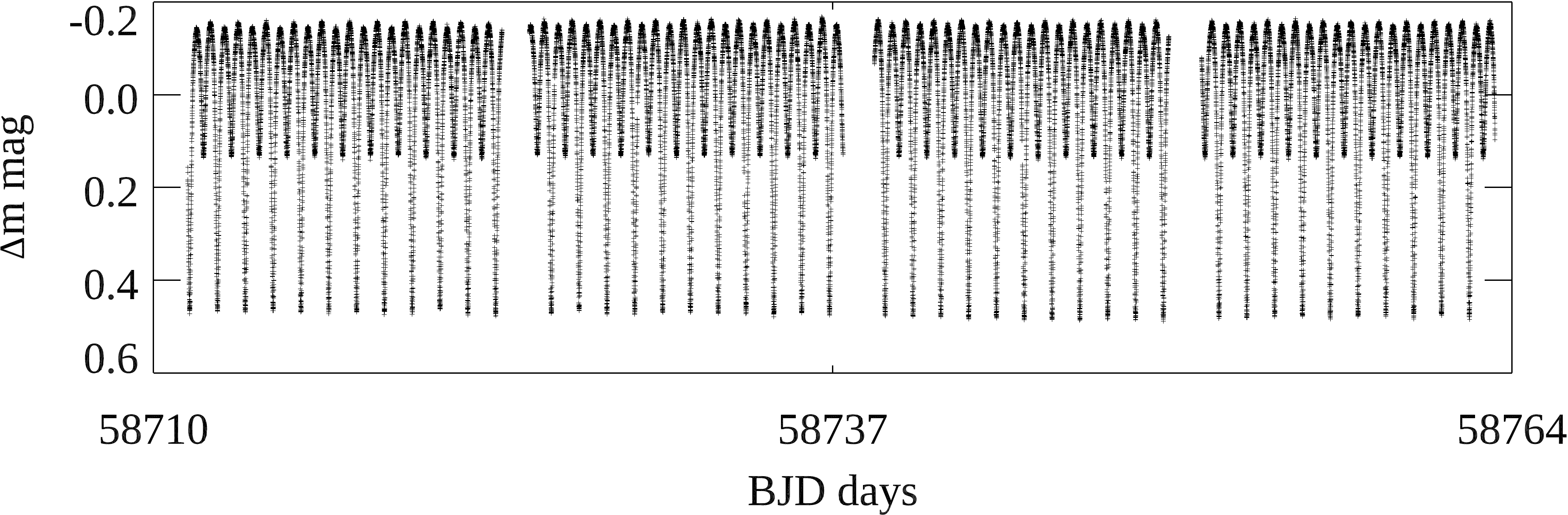}  	
\includegraphics[width=0.9\linewidth,angle=0]{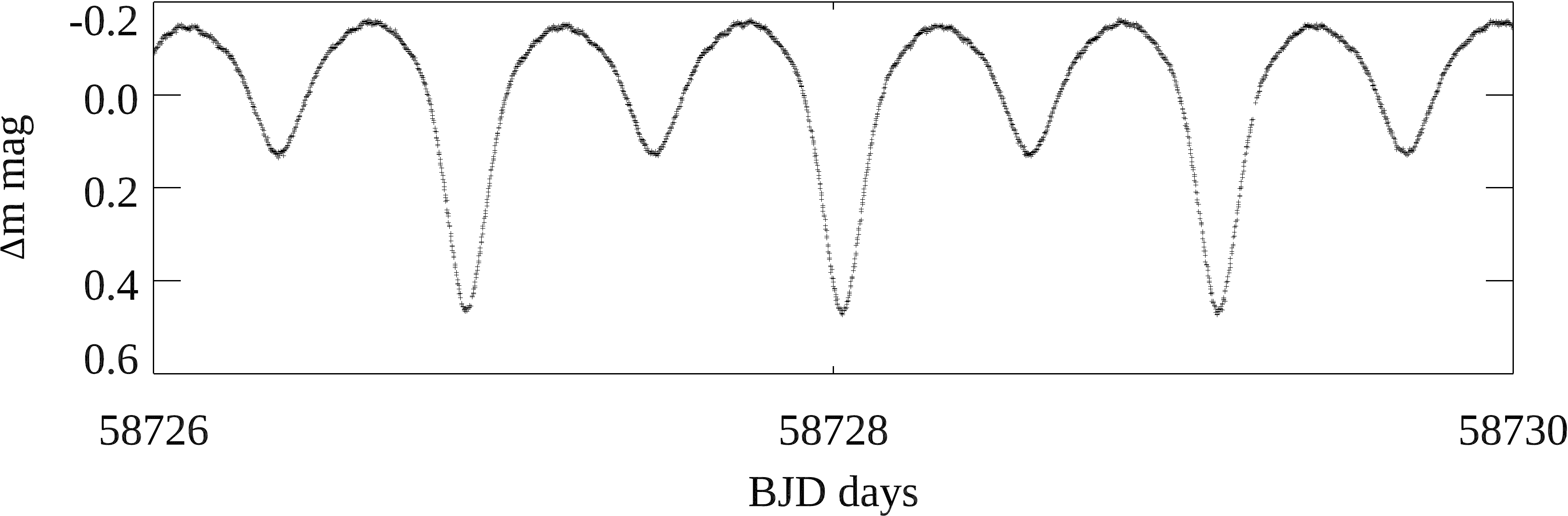} 	
\caption{Top: The full Sectors 15-16 light curve of TIC~63328020 showing the orbital variations. Bottom: A section of the light curve where close inspection shows the pulsations, particularly at orbital quadrature. The zero point of the ordinate scale is the mean.}
\label{fig:lc1}
\end{figure*}

Fig.~\ref{fig:lc1} shows the {\em TESS} photometry for TIC~63328020.  The top panel displays the full Sectors 15 and 16 light curve, where the eclipses and ellipsoidal light variations are obvious. The bottom panel shows a short segment of the light curve where more details can be seen.  Importantly, a careful look reveals that the pulsational variations are largest on the ELV humps where the star is brightest. That, of course, is at orbital quadrature, so this is distinctly different from the first two single-sided pulsators, HD~74423 and CO~Cam. The frequency analysis below bears out this first impression.  First, we look at the light curve, which shows the ellipsoidal orbital variations clearly, and the amplitude modulation of the pulsations on careful inspection.

\subsection{The orbital frequency}

We analysed the data using the frequency analysis package \mbox{{\sc period04}} \citep{2005CoAst.146...53L}\footnote{https://www.univie.ac.at/tops/Period04/}, a Discrete Fourier Transform program \citep{1985MNRAS.213..773K} to produce amplitude spectra, and a combination of linear and nonlinear least-squares fitting to optimise frequency, amplitude and phase. The derived orbital frequency is $\nu_{\rm orb} = 0.9043640 \pm 0.0000007$\,d$^{-1}$ ($P_{\rm orb} = 1.1057495 \pm 0.0000008$\,d), where variance from the pulsations and from some low frequency artefacts have been filtered for a better estimate of the uncertainty in the frequency. A 50-harmonic fit by least-squares was done to see how this frequency fits the data, and to show that the pulsation frequencies are not orbital harmonics. In practice, the harmonics at frequencies higher than $20 \times \nu_{\rm orb}$ do not have statistically significant amplitudes. In particular, the 20th orbital harmonic has S/N = 7 in amplitude. All higher harmonics have S/N $<3$ in amplitude.

After pre-whitening the data by the 50-harmonic fit, some low frequency artefacts from the data reduction were removed with a high-pass filter, which was a simple consecutive pre-whitening of low frequency peaks extracted by Fourier analysis until the noise level was reached in the frequency range $0 - 6$\,d$^{-1}$. This was done to study the pulsations with white noise for the purpose of estimating the uncertainties. 

\subsection{The pulsation}

TIC~63328020 pulsates principally in a single mode at a frequency of $\nu_1 = 21.09533 \pm 0.00014$\,d$^{-1}$, typical of $\delta$~Sct stars. Because this oblique nonradial pulsation mode is observed from changing aspect with the orbit of the star and its synchronous rotation, amplitude and phase modulation of the pulsation generate a frequency septuplet\footnote{Because this oblique nonradial pulsation mode is a distorted dipole mode observed from changing aspect with the orbit of the star and its synchronous rotation, amplitude and phase modulation of the pulsation generate a frequency septuplet (a pure dipole mode would generate a frequency triplet).}. There is also a harmonic at $2\nu_1$ that generates a quintuplet. These are typical of oblique pulsators. There are also two low-amplitude frequencies at 10.502\,d$^{-1}$ and 11.406\,d$^{-1}$ that are separated by the orbital frequency. It is likely that one of these is a mode frequency and the other is part of a frequency multiplet from oblique pulsation where the signal-to-noise is too low to detect the other multiplet components. Whichever of these two frequencies is the mode frequency, it is close to, but is not,  a sub-harmonic of the principal mode frequency.  These peaks at $10.502\,{\rm d}^{-1}$ and $11.406\,{\rm d}^{-1}$ have amplitude signal-to-noise ratios of 5 and 6, respectively, which are too low for further discussion of these frequencies here.

\begin{figure*}
\centering
\includegraphics[width=0.7\linewidth,angle=0]{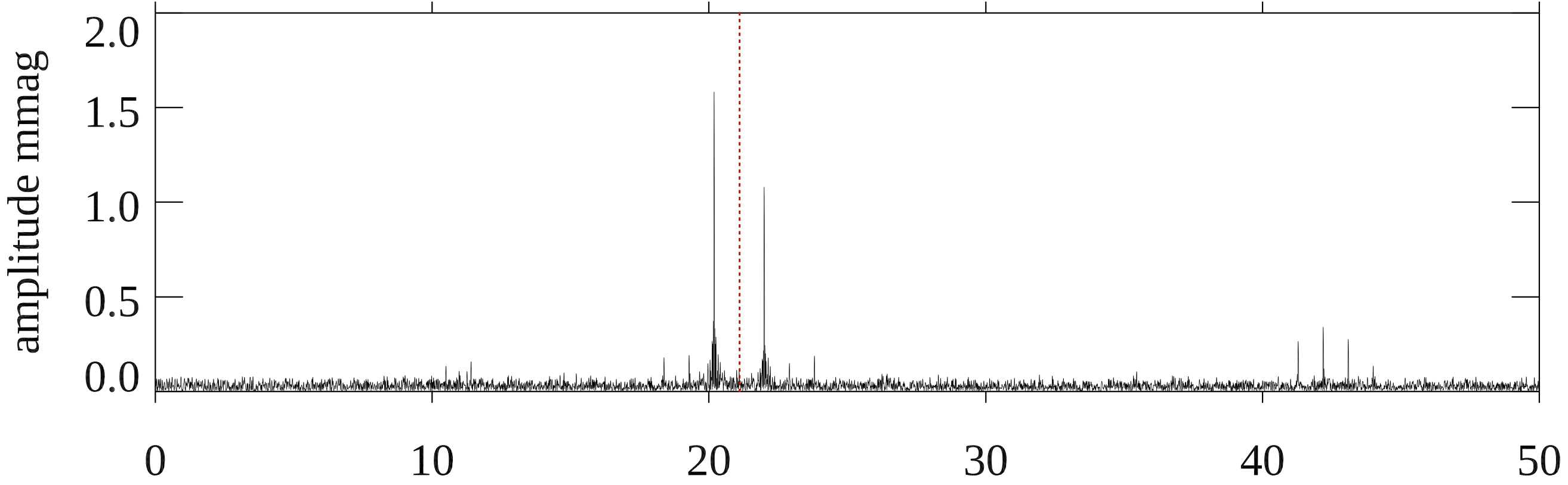} 	\vglue 0.1cm \hglue0.03cm
\includegraphics[width=0.7\linewidth,angle=0]{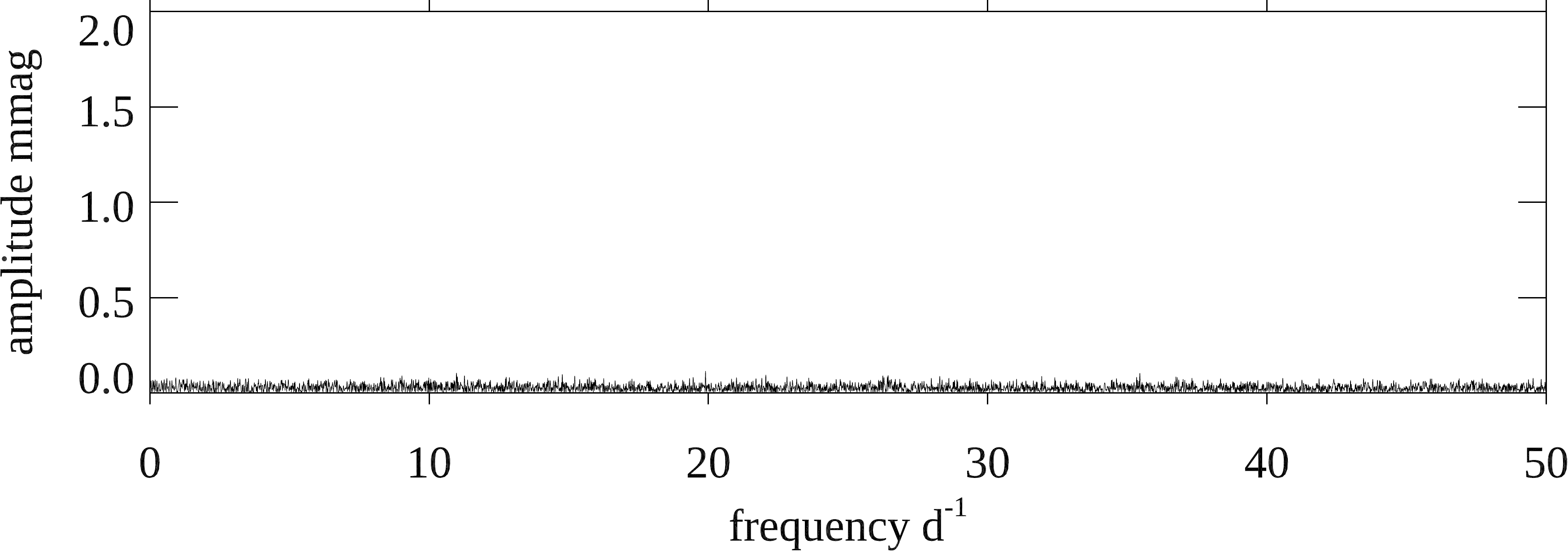} 	
\caption{Top: The amplitude spectrum of the high pass data. There is a frequency septuplet centred on $\nu_1 = 21.09533$ c/d, and a quintuplet centred on $2 \nu_1$. The reason there are two high amplitude first sidelobes about an almost zero amplitude mode pulsation frequency (marked by the vertical dotted red line) is that we are seeing this pulsation inclined by $90^\circ$ to the orbital axis, which itself is near to $i = 90^\circ$, thus giving two pulsation amplitude maxima per orbit. }
\label{fig:ft1}
\end{figure*}  

The top panel in Fig.\,\ref{fig:ft1} shows the amplitude spectrum for the high-pass filtered data. An inspection shows that there is a septuplet centred on $\nu_1 = 21.09533$\,d$^{-1}$ and a quintuplet centred on $2\nu_1 = 42.19065$\,d$^{-1}$. The detected frequencies are listed in Table \ref{table:freqs}.  Because the orbital inclination is close to $90^\circ$, and the tidal pulsation axis is inclined $90^\circ$ to that, the pulsation shows amplitude maxima twice per orbit, as will be seen in the next subsection. That then generates two principal peaks in the amplitude spectrum at $\nu_1 - \nu_{\rm orb}$  and $\nu_1 + \nu_{\rm orb}$, along with the other sidelobes. We determined the frequencies in the multiplets by a combination of linear and nonlinear least-squares fitting, and determined that all separations in the multiplets are equal to the orbital frequency $\nu_{\rm orb} = 0.9043640 \pm 0.0000007$\,d$^{-1}$ determined in the last section within 1.5$\sigma$. We took the average of those two highest amplitude sidelobe frequencies to determine the value of $\nu_1$, and generated the frequency multiplets from that.  Thus we only give frequency uncertainties on these two highest amplitude frequencies, i.e., those for the first orbital sidelobes $\nu_1 - \nu_{\rm orb}$ and $\nu_1 + \nu_{\rm orb}$, which are $\pm 0.00015$\,d$^{-1}$ and $\pm 0.00022$\,d$^{-1}$, respectively.

We then forced a frequency septuplet for $\nu_1$ and a quintuplet for $2\nu_1$, all split by exactly the orbital frequency,  $\nu_{\rm orb} = 0.9043640 \pm 0.0000007$\,d$^{-1}$. The reason for choosing this exact splitting is that for oblique pulsators it is instructive to examine the pulsation phases, and those are inextricably coupled to the frequencies, as can be seen by examining the function we fitted, ($\cos 2 \pi f (t - t_o) + \phi$). Pre-whitening by that solution (and including in the fit the two low-frequencies 10.502\,d$^{-1}$ and 11.406\,d$^{-1}$) leads to no variance in the data above noise, as can be seen in the bottom panel of Fig.\,\ref{fig:ft1}. This shows that exact splitting by the orbital frequency about $\nu_1$ and 2$\nu_1$ fits the data, and that there are no other detectable pulsations. This star has one principal oblique pulsating mode. 

Finally, we chose a time zero point, $t_0$, such that the pulsation phases were equal for the two dominant peaks seen in the top panel of  Fig.\,\ref{fig:ft1}. This effectively choses a time of pulsation amplitude maximum with the orbit and rotation of the star, thus provides the orbital phase for the time of pulsation maximum. Table~\ref{table:freqs} shows a least-squares fit of the determined frequencies. 

\begin{table}
\centering
\caption{A least squares fit of the two low frequencies, the frequency septuplet for $\nu_1$ and the frequency quintuplet for $2\nu_1$.  The zero point for the phases, $t_0 = {\rm BJD}\,2458737.14936$,  has been chosen to be a time when the two first orbital sidelobes of $\nu_1$ have equal phase.} 
\begin{tabular}{rrcr}
\hline
&\multicolumn{1}{c}{frequency} & \multicolumn{1}{c}{amplitude} &   
\multicolumn{1}{c}{phase}  \\
&\multicolumn{1}{c}{d$^{-1}$} & \multicolumn{1}{c}{mmag} &   
\multicolumn{1}{c}{radians}   \\
& & \multicolumn{1}{c}{$\pm 0.023$} &   
   \\
\hline
 $\nu_{\rm low} - \nu_{\rm orb}$& 10.50197  &  0.133  &  $-0.708 \pm 0.175 $ \\
 $\nu_{\rm low}$& 11.40633  &  0.152  &  $1.129 \pm 0.152 $ \\
 \hline 
 $\nu_1 - 3\nu_{\rm orb}$& 18.38223  &  0.181  &  $1.350 \pm 0.128 $ \\
 $\nu_1 - 2\nu_{\rm orb}$& 19.28660  &  0.192  &  $2.859 \pm 0.121 $ \\
 $\nu_1 - \phantom{1}\nu_{\rm orb}$& 20.19096  &  1.585  &  $0.719 \pm 0.015 $ \\
 $\nu_1$ & 21.09533  &  0.110  &  $0.304 \pm 0.211 $ \\
 $\nu_1 + \phantom{1}\nu_{\rm orb}$& 21.99969  &  1.080  &  $0.719 \pm 0.022 $ \\
$\nu_1 + 2\nu_{\rm orb}$ & 22.90405  &  0.154  &  $-0.419 \pm 0.151 $ \\
$\nu_1 + 3\nu_{\rm orb}$ & 23.80842  &  0.185  &  $1.296 \pm 0.125 $ \\
\hline
 $2\nu_1 - 2\nu_{\rm orb}$& 40.38193  &  0.049  &  $-2.882 \pm 0.479 $ \\
 $2\nu_1 - \phantom{1}\nu_{\rm orb}$& 41.28629  &  0.266  &  $1.608 \pm 0.087 $ \\
$2\nu_1$ & 42.19065  &  0.341  &  $0.057 \pm 0.068 $ \\
 $2\nu_1 + \phantom{1}\nu_{\rm orb}$& 43.09502  &  0.277  &  $-1.652 \pm 0.084 $ \\
$2\nu_1 + 2\nu_{\rm orb}$ & 43.99938  &  0.133  &  $-3.130 \pm 0.175 $ \\
 \hline
\hline
\end{tabular}
\label{table:freqs}
\end{table}

\subsection{Pulsation as a function of orbital phase}

While the frequencies, amplitudes and phases determined by Fourier analysis and least-squares fitting in the last subsection contain the information to study the oblique pulsation, it is instructive and easier to see how this pulsation varies with orbital aspect by plotting the pulsation amplitude and phase as a function of orbital (rotational) phase. To do this we fitted the pulsation frequency, $\nu_1$, and its harmonic, $2\nu_1$, to chunks of the data that are $P_{\rm orb}/10$ in duration. It is immediately obvious from doing this that pulsation amplitude maximum occurs in quadrature to the orbital eclipses. This is the signature of an oblique pulsation in a dipole sectoral mode. 

To show this, we have set the time zero point $\frac{1}{4}$ of an orbital period prior to pulsation maximum, as determined from the fitting in the last subsection. That zero point is  $t_0 = {\rm BJD}\,2458736.87292$. We emphasise that this time has been chosen with reference to the pulsation amplitude maximum, hence is independent of the determination of the time of primary eclipse from the study of the orbital variations.  

\begin{figure}
\centering
\includegraphics[width=1.0\linewidth,angle=0]{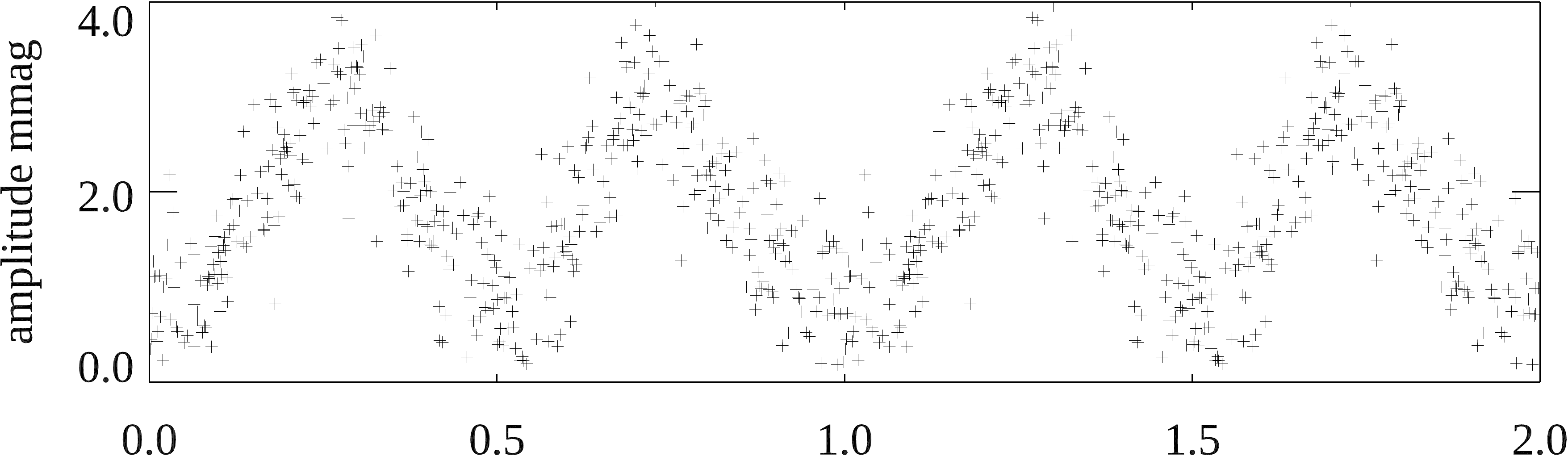} 	
\includegraphics[width=1.0\linewidth,angle=0]{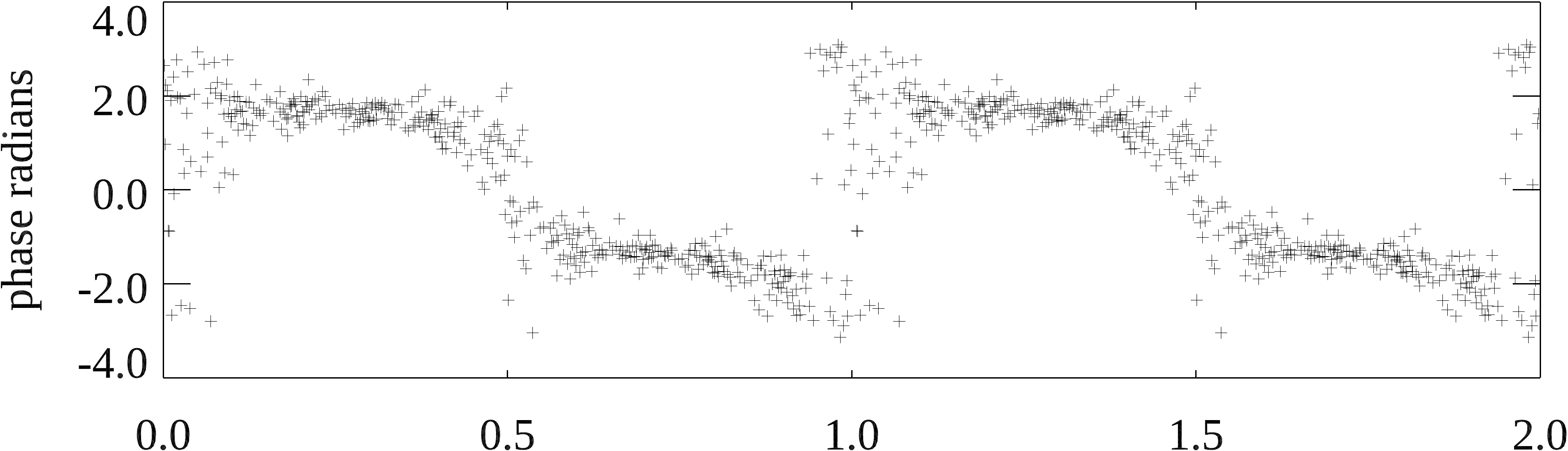} 
\includegraphics[width=1.0\linewidth,angle=0]{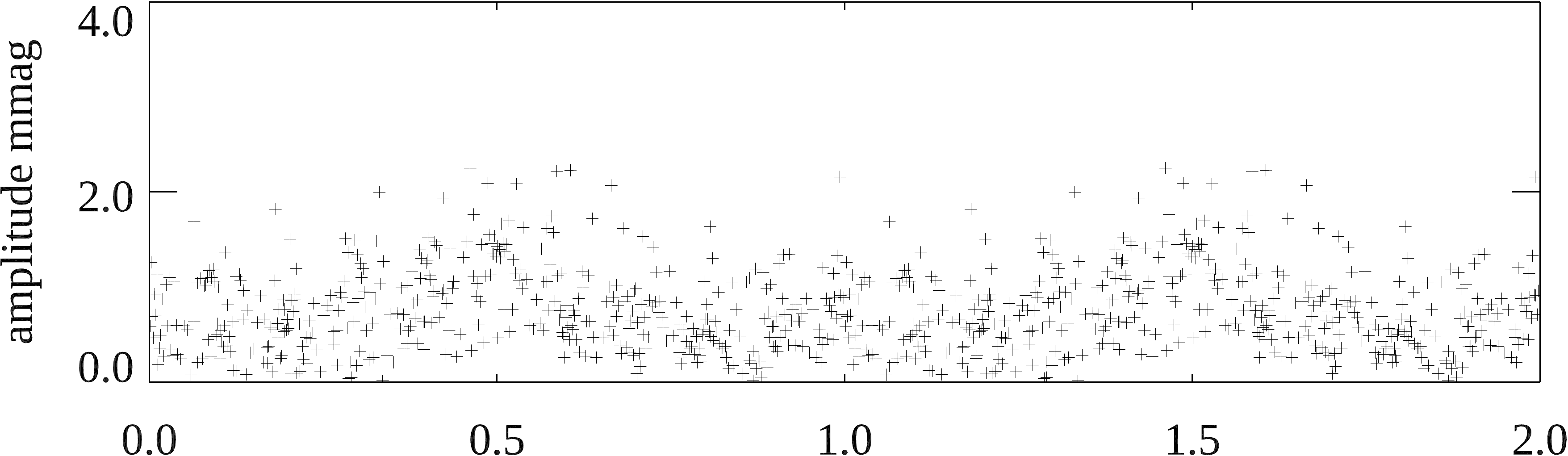} 	
\includegraphics[width=1.0\linewidth,angle=0]{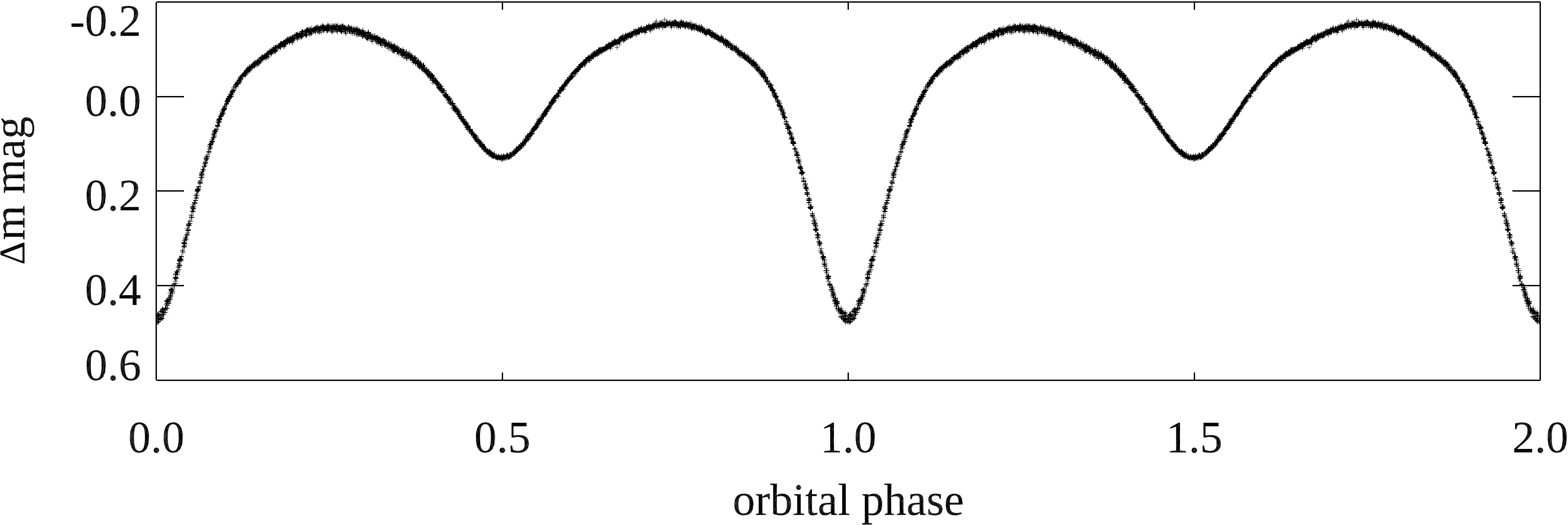} 
\caption{$\nu_1$: Top and second panels: The pulsation amplitude and phase variation as a function of orbital phase taking $\nu_1$ to be the pulsation frequency. The zero point in time,  $t_0 = {\rm BJD}\,2458736.87292$, has been set to be $\frac{1}{4}$ of an orbital period before the time of pulsation maximum, which was determined by choosing the time when the phases of the first orbital sidelobes are equal.  Third panel: The pulsation amplitude as a function of orbital phase for the second harmonic 2$\nu_1$. The phase diagram for this frequency is not shown, since it is uninformative because the low amplitude results in high scatter in the phase determinations. The second harmonic peaks when the L$_3$ point is closest to the observer. Bottom: the orbital light variations as a function of orbital phase for comparison. For this panel the data were binned by a factor of 10.  It can be seen that orbital light maximum coincides with pulsation maximum, and that the pulsation amplitude is very similar at orbital phases 0.0 and 0.5. This is currently unique.}
\label{fig:phamp_nu1}
\end{figure}

Fig.\,\ref{fig:phamp_nu1} shows the results. The mode is a sectoral dipole mode with $\ell = 1, |m| = 1$. Amplitude maximum coincides with orbital quadrature, and the phase reverses by $\pi$\,rad at the times of the eclipses when the line of sight aligns with the tidal axis. This is new and currently unique among tidally tilted pulsators. Because the mode is sectoral, it has a symmetry with respect to the tidal distortion such that the star is not strongly ``single-sided''. The third panel of Fig.\,\ref{fig:phamp_nu1} shows that the harmonic distortion of the mode is strongest during secondary eclipse when the L$_3$ side of the pulsating star is closest to the observer, thus the L$_1$ and L$_3$ sides of the pulsator do differ and the star is mildly a ``single-sided pulsator''.

\subsection{An identification constraint on the p mode}

The standard simple relation for a toy model pulsator of $P \sqrt{\frac{\rho}{\rho_\odot}} = Q$ can be rewritten in terms of observables as \\

$\log Q = \log P + \frac{1}{2} \log g + \frac{1}{10} M_{\rm bol} +\log T_{\rm eff} -6.454 $,
\\

\noindent where $Q$ is a pulsation ``constant'' that can be compared with models, $P$ is the pulsation period in days and $\log g$ is in cgs units. Taking $T_{\rm eff} \simeq 8000$\,K, $\log g \simeq 3.8$ (Sect.~\ref{sec:mcmc}), and  $M_{\rm bol} \simeq M_V \simeq 1.86$\footnote{The bolometric correction near F0 is zero (see Table 2 of \citealt{mortonadams68}).} from the Gaia parallax and $V$ magnitude then gives for $\nu_1$ a value of Q = 0.016, indicative of radial overtone around $n = 3-4$. The same calculation for the low frequency peak at 11.406\,d$^{-1}$ gives $Q = 0.021$, suggesting a second radial overtone mode. 

\section{System Properties of TIC~63328020}
\label{properties}

\subsection{Archival Data}
\label{sec:archive}

\subsubsection{Magnitudes and Gaia Results}
\label{magnitudes}

We have collected a set of archival magnitudes for TIC~63328020 and report these values in Table~\ref{tbl:mags}.  Additionally, we list the Gaia information about this object in Table \ref{tbl:mags}.  There is a fainter ($G = 19.9$) neighbour star some 2.84$^{\prime\prime}$ away, but there is insufficient Gaia information to tell whether that star is physically associated with TIC~63328020.

\begin{table}
\centering
\caption{Properties of the TIC~63328020 System}
\begin{tabular}{lc}
\hline
\hline
Parameter & Value   \\
\hline
RA (J2000) (h m s)& 21:20:14.41  \\  
Dec (J2000) ($^\circ \ ^\prime \ ^{\prime\prime}$) &  51:23:41.04 \\ 
$T$$^a$ & $11.553 \pm 0.014$ \\
$G$$^b$ & $11.967 \pm 0.009$  \\
$G_{\rm BP}$$^b$ & $12.267 \pm 0.026$  \\
$G_{\rm RP}$$^b$ & $11.462 \pm 0.024$  \\
$B^a$ & $12.619 \pm 0.228$ \\
$V^a$ & $11.965 \pm 0.114$ \\
$J^c$ & $10.991 \pm 0.020$   \\
$H^c$ & $10.829 \pm 0.019$  \\
$K^c$ & $10.744 \pm 0.018$ \\
W1$^d$ & $10.451 \pm 0.024$ \\
W2$^d$ & $10.476 \pm 0.022$  \\
W3$^d$ & $10.232 \pm 0.070$  \\
W4$^d$ & $> 9.07$  \\
$R$ (${\rm R}_\odot$)$^b$ & $3.6^{+0.2}_{-0.6}$  \\
$L$ (${\rm L}_\odot$)$^b$ & $13.5 \pm 0.8 $  \\
Orbital Period (d)$^e$ & $1.10575$ \\
$K_1$ (km~s$^{-1}$)$^e$ & $87.5 \pm 4.5$\\
$\gamma$ (km~s$^{-1}$)$^e$ & $-8.8 \pm 3.6$ \\
Distance (pc)$^b$ & $ 1054 \pm 39$  \\   
$\mu_\alpha$ (mas ~${\rm yr}^{-1}$)$^b$ & $+2.51 \pm 0.05$   \\ 
$\mu_\delta$ (mas ~${\rm yr}^{-1}$)$^b$ &  $+1.49 \pm 0.05$   \\ 
\hline
\label{tbl:mags}  
\end{tabular}

{\bf Notes.}  (a) ExoFOP (exofop.ipac.caltech.edu/tess/index.php).  (b) Gaia DR2 (\citealt{2018A&A...616A...2L}; 
\citealt{2018A&A...616A...9L}; \citealt{2018A&A...616A...1G}).  (c) 2MASS catalog \citep{2006AJ....131.1163S}.  (d) WISE point source catalog \citep{2013yCat.2328....0C}.  (e) This work; see Table \ref{tbl:porb} for details on the orbital period and Table \ref{tbl:RVs} and Fig.~\ref{fig:RVs} for the RV results.
\end{table}

\subsubsection{Spectral Energy Distribution}

\begin{figure}
\centering
\includegraphics[width=1.0\columnwidth]{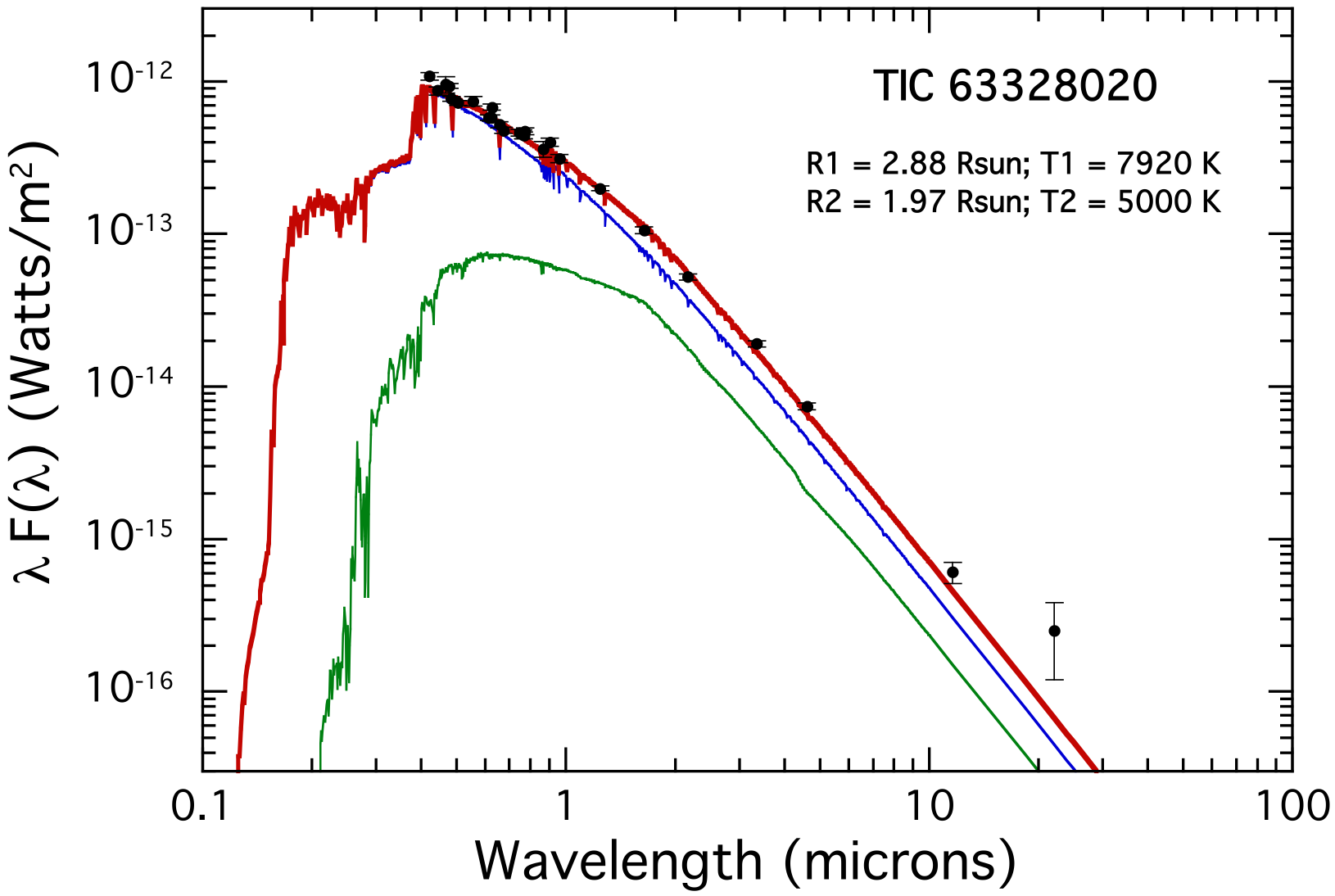}	
\caption{SED data and model for TIC~63328020 spanning the blue to 20 micron region for the non-coeval case described in Section~\ref{sec:mcmc_no_coeval}.  The continuous green, blue and red curves represent the contributions of the secondary, the primary, and the total system flux, respectively. }
\label{fig:sed}
\end{figure}

The spectral energy distribution (SED) points for this object are plotted in flux units in Fig.~\ref{fig:sed}, and many of them are reported as magnitudes in Table \ref{tbl:mags}.  The SED points have all been corrected for interstellar extinction at a level\footnote{(http://argonaut.skymaps.info/query)}  of $E(g-r) \simeq 0.30 \pm 0.03$ \citep{2018MNRAS.478..651G,2019ApJ...887...93G}, which we take to mean $A_V \simeq 1$. We then use the wavelength dependence of extinction given by \citet{cardelli89}, in particular, the fitting formulae in their Eqn.~(2) and (3) for $\langle A(\lambda)/A(V) \rangle$. Also shown on the figure are fitted curves to the SED based on the contributions from both stars in the binary.  These will be discussed in Section~\ref{sec:mcmc_no_coeval}.

\subsubsection{Archival Photometric Data}

\begin{figure}
\centering
\includegraphics[width=1.0\columnwidth]{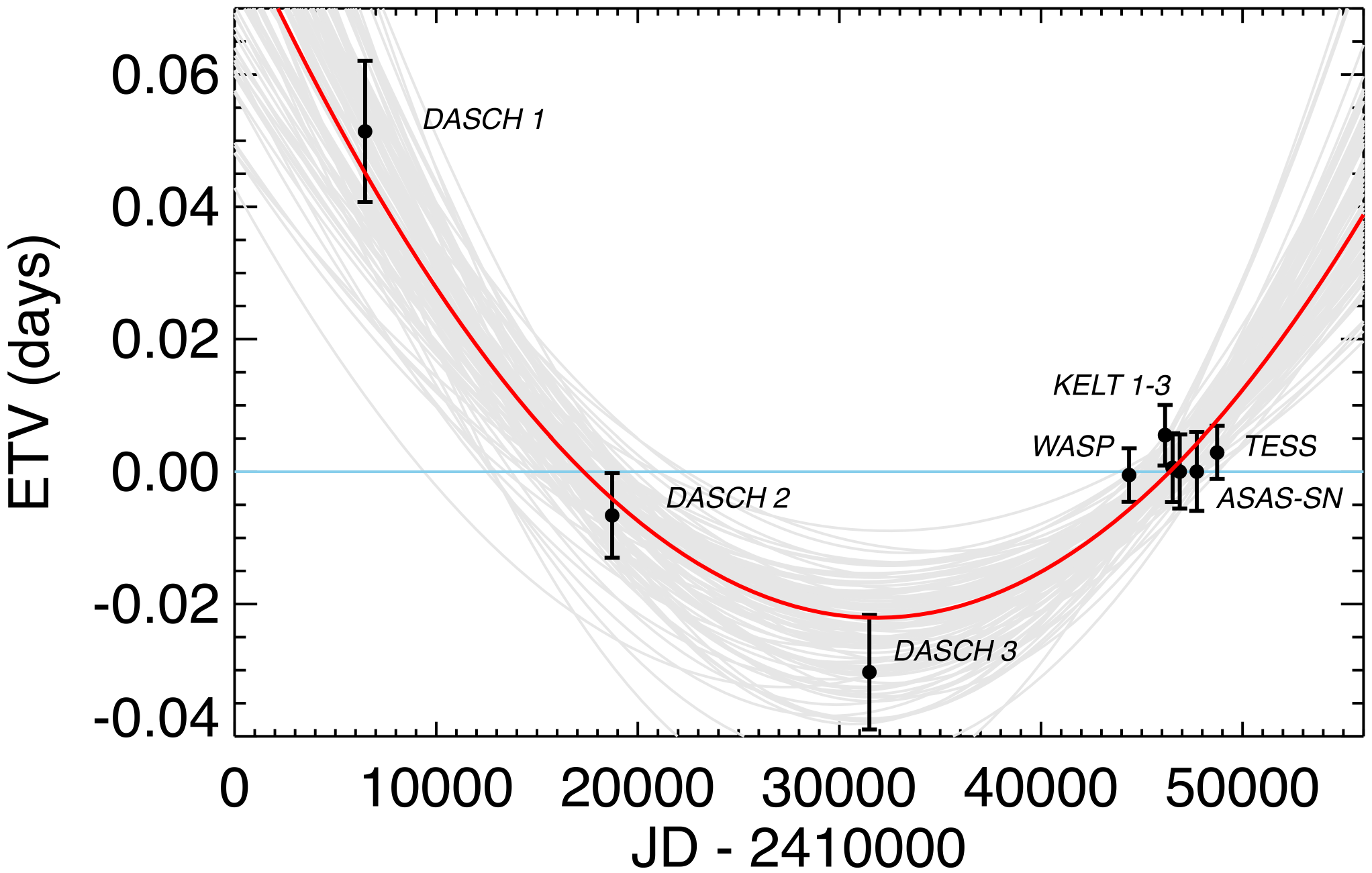}
\caption{Long-term eclipse timing variations for TIC~63328020.  The data sets used are marked with labels and include {\em TESS}, ASAS-SN, WASP, KELT, and DASCH (see Table \ref{tbl:ETVs} for references).  The latter two data sets have each been divided into three subsets of 30 years and $\sim$10 months, respectively.  The time intervals covered by each data set and the references are given in Table \ref{tbl:ETVs}. The error bars plotted here and used in the long-term fit are taken from Eqn.\,(\ref{eqn:sigma2}) with $\sigma_J$ found from the fit to be 0.0057 d (see text and Eqn.\,\ref{eqn:sigma1}).  We did this so as not to allow the tiny error bars of the modern-epoch points to totally dominate the determination of the long-term period or its derivative. The red curve is the best fit to the linear plus quadratic terms. See Table \ref{tbl:porb} for results. The multiple curves in faint grey are 100 random draws from the posteriors of the MCMC fit. }
\label{fig:etv_phot}
\end{figure}  

\begin{table}
\centering
\caption{Eclipse Timing Variation Data for TIC~63328020}
\begin{tabular}{lccc}
\hline
\hline
Source & Start Date$^a$ & End Data$^a$ & ETV$^b$ \\
\hline
{\em TESS}$^c$ & 58711 & 58763 & $+0.0029 \pm 0.0003$ \\ 
ASAS-SN$^d$ & 57106 & 58348 & $+0.00002 \pm 0.0044$ \\ 
KELT 3$^e$ &  56744  & 57021  & $+0.00002 \pm 0.0039$ \\  
KELT 2$^e$ &  56387  & 56668  & $+0.0006 \pm 0.0033$ \\  
KELT 1$^e$ &  56007  & 56303  & $+0.0055 \pm 0.0022$ \\      
WASP$^f$ &  54070  & 54670 & $-0.0005 \pm 0.0007$ \\                             
DASCH 3$^g$ & 35119 & 47857 & $-0.0303 \pm 0.0077$ \\  
DASCH 2$^g$ & 22335 & 35119 & $-0.0066 \pm 0.0050$ \\  
DASCH 1$^g$ & 10606 & 22335 & $+0.0514 \pm 0.0099$ \\  
\hline
\label{tbl:ETVs}  
\end{tabular}

{\bf Notes.}  (a) HJD-2\,400\,000. (b) The fold period is 1.105\,769\,8 days with an epoch of HJD 2\,444\,000.2795. The ETV value is expressed in days. (c) This work. (d) \citet{shappee14} and \citet{kochanek17}.  (e) \citet{2007PASP..119..923P} and \citet{2012PASP..124..230P}.  (f) \citet{colliercameron06}; \citet{pollacco06}. (g) \citet{grindlay09}.  
\end{table}

\begin{table}
\centering
\caption{Orbital Period Determinations for TIC~633280200}
\begin{tabular}{lcc}
\hline
\hline
Parameter & Value & Uncertainty  \\
\hline
& {\em TESS} Only &   \\
\hline
$P_{\rm orb}$$^a$ [days] &1.105\,749  & 0.000\,001 \\
$P_{\rm orb}$$^b$ [days] &1.105\,9~~~~ & 0.000\,2~~~~ \\              
$P_{\rm orb}$$^c$ [days] & 1.105\,751 & 0.000\,001  \\  
$P_{\rm orb}$$^d$ [days] & 1.105\,754  & 0.000\,001 \\
$P_{\rm orb}/\dot P_{\rm orb}$$^e$ [years] &  $-$3050  & 500 \\  
\hline
& Long-Term ETV Study & \\
\hline
$P_{\rm orb}$$^f$ [days] & 1.105\,769\,8 & 0.000\,000\,3  \\  
$P_{\rm orb}$$^g$ [days] & 1.105\,770\,3 & 0.000\,000\,3 \\ 
$P_{\rm orb}/\dot P_{\rm orb}$$^h$ [years] &  $+12.4 \times 10^6$ &   $3.0 \times 10^6$ \\ 
Jitter Noise, $\sigma_J$$^i$ [days] & 0.0057 & $^{+0.0031}_{-0.0018}$ \\ 
\hline
\label{tbl:porb}  
\end{tabular}

{\bf Notes.}  (a) Based on the frequency analysis. (b) Spacing between the $\nu_1$ pulsation sextuplet. The corresponding epoch time is JD 2\,458\,736.8729.  (c) Eclipse timing analysis (`ETV') assuming no period changes. The reference fold epoch is JD 2\,458\,736.8720. (d) ETV analysis allowing for $\dot P_{\rm orb}$ . (e) $P_{\rm orb}/\dot P_{\rm orb}$ from the ETV analysis.  (f) From the long-term photometric ETV analysis assuming $P_{\rm orb}$ is a constant. The fold epoch time is JD 2\,444\,000.2795. (g) From the long-term photometric ETV analysis allowing for $\dot P_{\rm orb}$.  (h) $P_{\rm orb}/\dot P_{\rm orb}$ from the long-term photometric ETV analysis (i) See Eqn.\,\ref{eqn:sigma2} for definition. 
\end{table}

In addition to the new {\em TESS} photometry on TIC~63328020, we have utilized archival photometric data from ASAS-SN, WASP, KELT, and DASCH (for references see Table \ref{tbl:ETVs}) to establish the long-term orbital ephemeris for this binary.  The time intervals for the various data sets, and references to the archival data are given in Table \ref{tbl:ETVs}.  The DASCH data cover more than a century, but only about 1100 scanned plates for this object were available; we divided these up into three roughly 30-yr long intervals which are denoted ``1", ``2" and ``3". The same was done for the KELT data which spanned three observing seasons and was divided into three $\sim$10-month segments.  For each data set we derived a time of mid-primary eclipse from a fold of the data about a common reference period and epoch (see Table \ref{tbl:ETVs}). 

From this set of archival photometry we derived a long-term ETV curve for this object which is plotted in Fig.~\ref{fig:etv_phot}.  While the long-term orbital period is well defined to a about a part per million with $P_{\rm orb} = 1.105\,769\,8(3)$ d, it is also apparent that there are significant non-linear ETVs present.  At the moment, there is insufficient information to quantify whether these are due to orbital motion induced by a distant companion or some other effect causing jitter in $P_{\rm orb}$ (e.g., \citealt{applegate92}).  To get a handle on the long-term trend in the orbital period, we modelled the measured ETVs as a quadratic function. We fitted simultaneously for a ``jitter'' term that we added in quadrature to the measured statistical uncertainties.  The jitter term models an independent noise term in our ETV measurements that is not captured by our formal uncertainties (perhaps a systematic uncertainty due to the way we measured the ETVs). Our log likelihood function was: 
\begin{equation}
\log{L} = -\sum_i \left[\frac{(y_i-m_i)^2}{2\sigma_i^2} + \log{\sigma_i} \right]
\label{eqn:sigma1}
\end{equation}
where $y_i$ are the measured ETVs, $m_i$ are evaluations of the quadratic model, and 
\begin{equation}
\sigma_i = \sqrt{\sigma_{0,i}^2 + \sigma_J^2} ~~,
\label{eqn:sigma2}
\end{equation}
where $\sigma_{0,i}$ are the formal uncertainties on the ETVs and $\sigma_J$ is the extra uncertainty added in quadrature.  The four free parameters, constant, linear, quadratic and $\sigma_J$ were found via an MCMC fit (see, e.g.,\,\citealt{2005AJ....129.1706F}).

 In Table \ref{tbl:porb} we summarize all the information that we have about the orbital period, and its derivative, derived in several different ways from the various available data sets.

\subsection{Spectral Measurements}
\label{sec:spec}

TIC~63328020 was observed with the Intermediate Dispersion Spectrograph (IDS) on the 2.5-m Isaac Newton Telescope (INT) between 28 November and  1 December 2019.  The blue-sensitive EEV10 detector was used along with a $1^{\prime\prime}$ wide slit and the R1200B grating centred at 4000\,\AA{} for an unvignetted spectral coverage of $\sim$3600--4600\,\AA{} at a resolution of approximately 4500.  In total, eight exposures were acquired with integration times between 1200\,s and 1400\,s, the dates of which are shown in Table~\ref{tbl:RVs}.  The spectra were wavelength calibrated against arc frames, illuminated using Copper-Argon and Copper-Neon lamps, obtained immediately after each observation in order to avoid shifts due to flexure in the instrument.  Bias subtraction, wavelength calibration, sky subtraction and optimal extraction \citep[following the routine of][]{horne86} were performed using standard \textsc{starlink} routines.

\subsubsection{Radial Velocities}
\label{sec:RVs}

We extracted the $K$-velocity of the primary star via two different approaches.  In the first we simply fit the deep Ca II K line (at 3934\,\AA) and thereby estimated RVs with the corresponding uncertainties.  In the second, we did a cross-correlation analysis.  For the latter, we removed the spectrograph blaze function from the spectra by breaking the spectra into 4-nm wide bins, identifying the highest 10~per~cent of flux measurements within each bin (most of which are in the continuum), fitting a basis spline to the continuum points, and dividing the spectrum by the best-fit spline. We then measured radial velocities by cross-correlating each blaze-corrected spectrum with the highest signal-to-noise observation. The results are very similar to those obtained from the Ca II K line alone, but the uncertainties for the cross-correlation result are empirically somewhat smaller.  We list both sets of RVs in Table \ref{tbl:RVs}.  

We do not see any lines from the companion star.  We estimate that the luminosity of the secondary is $\lesssim$10~per~cent that of the primary.

\begin{table}
\centering
\caption{Radial Velocity Data for TIC~63328020}
\begin{tabular}{lccc}
\hline
\hline
Epoch (BJD) & RV (Ca II K)$^a$ & RV (CCF)$^b$ & RV (average)$^c$ \\
   ($-$2450000)  & km~s$^{-1}$   & km~s$^{-1}$  & km~s$^{-1}$  \\
8816.3148 & $70.9 \pm 4.1$ & $67.7 \pm 3.6$ & $69.1 \pm 3.4$ \\  
8816.4059 & $37.1 \pm 4.3$ & $30.7 \pm 3.6$ & $33.3 \pm 3.5$ \\
8817.3112 &  $78.0 \pm 3.7$  & $77.5 \pm 3.6$ & $77.8 \pm 3.3$ \\  
8817.3870 & $69.0 \pm 3.9$ & $70.1 \pm 3.6$ & $69.6 \pm 3.3$ \\ 
8818.3317 & $62.8 \pm 4.1$ & $67.0 \pm 3.6$ & $65.2 \pm 3.4$ \\ 
8818.4005 & $72.3 \pm 5.7$ & $77.0 \pm 3.6$ & $75.7 \pm 3.9$ \\ 
8819.3069 & $14.1 \pm 4.2$ & $12.0 \pm 3.6$ & $12.9 \pm 3.5$ \\ 
8819.3908 & $55.4 \pm 3.8$ & $58.5 \pm 3.6$ & $57.0 \pm 3.3$ \\ 
\hline
Epoch of $\phi_0$ & $K_1$$^d$ & $\gamma$$^d$ & cycles after {\em TESS} \\
           & km~s$^{-1}$   & km~s$^{-1}$      &   \\ 
\hline
8818.7046(42) & $87.0 \pm 5.2$ & $-8.2 \pm 4.2$ & $74.006 \pm 0.004$ \\
\hline
\label{tbl:RVs}  
\end{tabular}

{\bf Notes.}  (a) Based on measurements of the Ca II K line only. (b) Based on a cross correlation analysis of the spectrum over the region of 3700 to 4700\,\AA{}.  (c) Formal statistically weighted average of the two RV analyses. (d) Based on the average of the RVs, and a simple sinusoidal fit that does not take into account the Rossiter-McLaughlin effect (but see Section~\ref{sec:phoebe} for a more complete analysis).
\end{table} 

\begin{figure}
\centering
\includegraphics[width=1.0\columnwidth]{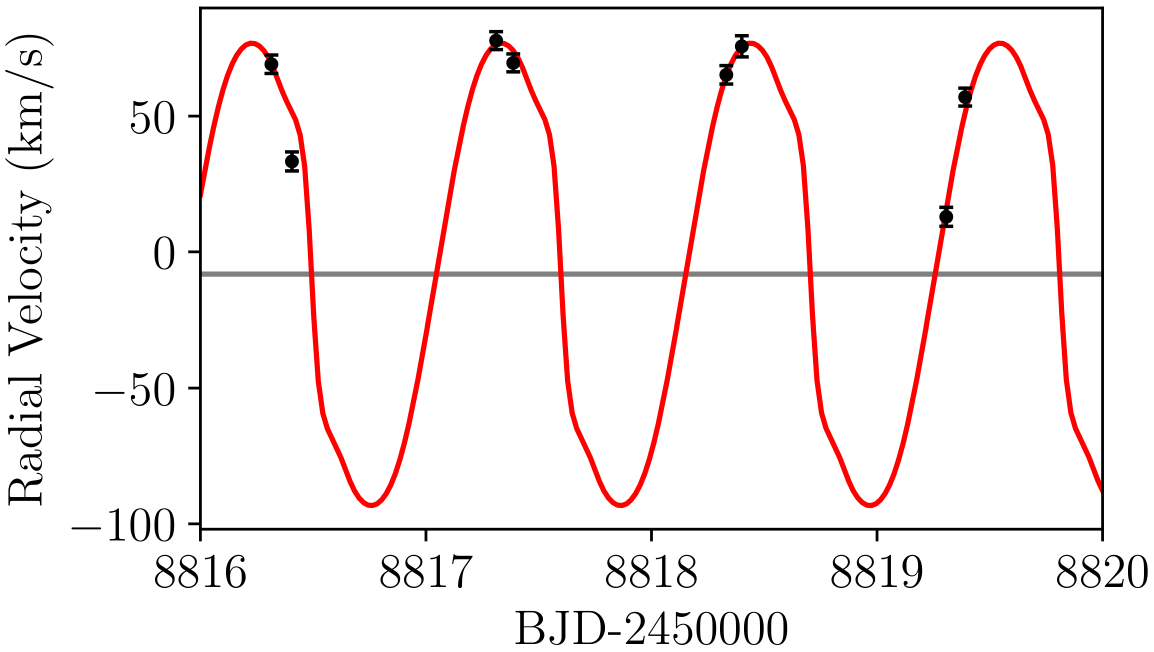}	
\caption{Radial velocity curve for the primary in TIC~63328020.  Because the orbital period is relatively close to a day, the orbital phase sampled over the four nights of observations did not cover the lower half of the RV curve.  The red curve is a model fit using {\tt Phoebe2}, yielding $89 \pm 6$ km/s (see Sect.~\ref{sec:phoebe}).}
\label{fig:RVs}
\end{figure}  

The RV data were taken over four consecutive nights, and since the orbital period is 1.1\,d, there was insufficient time for the orbital phases during the observations to drift more than $\sim$40\% of an orbital cycle.  Nonetheless, we were able to measure the $K$ velocity of the primary star to be $87.5 \pm 4.5$ km s$^{-1}$ (see Table \ref{tbl:RVs}). The radial velocities are plotted in Fig.~\ref{fig:RVs} along with a superposed model fit which is discussed in Section~\ref{sec:phoebe}.  

The RV orbital phase zero comes out to be $82.006 \pm 0.005$ orbital cycles after the {\em TESS} reference phase zero, and thus is consistent to within an uncertainty of $\sim$8 minutes. The RV orbital phase zero is also consistent with the orbital phase zero determined from the pulsation frequencies (see Fig.\,\ref{fig:phamp_nu1}), with the same uncertainty. This shows how precisely the pulsation axis coincides with the tidal axis. 

\subsubsection{Spectrally Determined Stellar Properties}
\label{sec:slellar_prop}

We also analyzed these same spectra to extract some basic properties of the primary star.  Because there is no sign of the cooler binary component in the available spectra, we simply assumed that the primary star dominates the spectral signatures.  The combined spectrum was used to determine the stellar atmospheric parameters ($T_{\rm eff}$, surface gravity $\log g$, metallicity Fe/H) and also the projected rotational velocity ($v \sin i$). During the spectral analysis, the Kurucz line list\footnote{kurucz.harvard.edu/linelists.html} was considered and also the ATLAS9 theoretical model atmospheres \citep{1993KurCD..13.....K} were generated with the SYNTHE code \citep{1981SAOSR.391.....K}. The synthetic spectra were compared with the combined spectrum to obtain the final atmospheric parameters with a $\chi^2$ minimization method. By using this approach, first we determined the $T_{\rm eff}$ value from the $H_{\gamma}$ line which is very sensitive to $T_{\rm eff}$. The $T_{\rm eff}$ value was searched over the range of $7000 - 8500$ K and for $T_{\rm eff}$\,$\leq$\,8000 K, $\log g$ was fixed to be 4.0 (cgs) because the hydrogen lines weakly depend on  $\log g$ for stars having $T_{\rm eff}$ values over 8000\,K \citep{2002A&A...395..601S}. The result was a derived $T_{\rm eff}$ of 7900\,$\pm$\,150 K. The comparison of the synthetic and the observed $H_{\gamma}$ line is shown in the lower panel of Fig.\,7. 

\begin{figure}
\centering
\includegraphics[width=1.0\columnwidth]{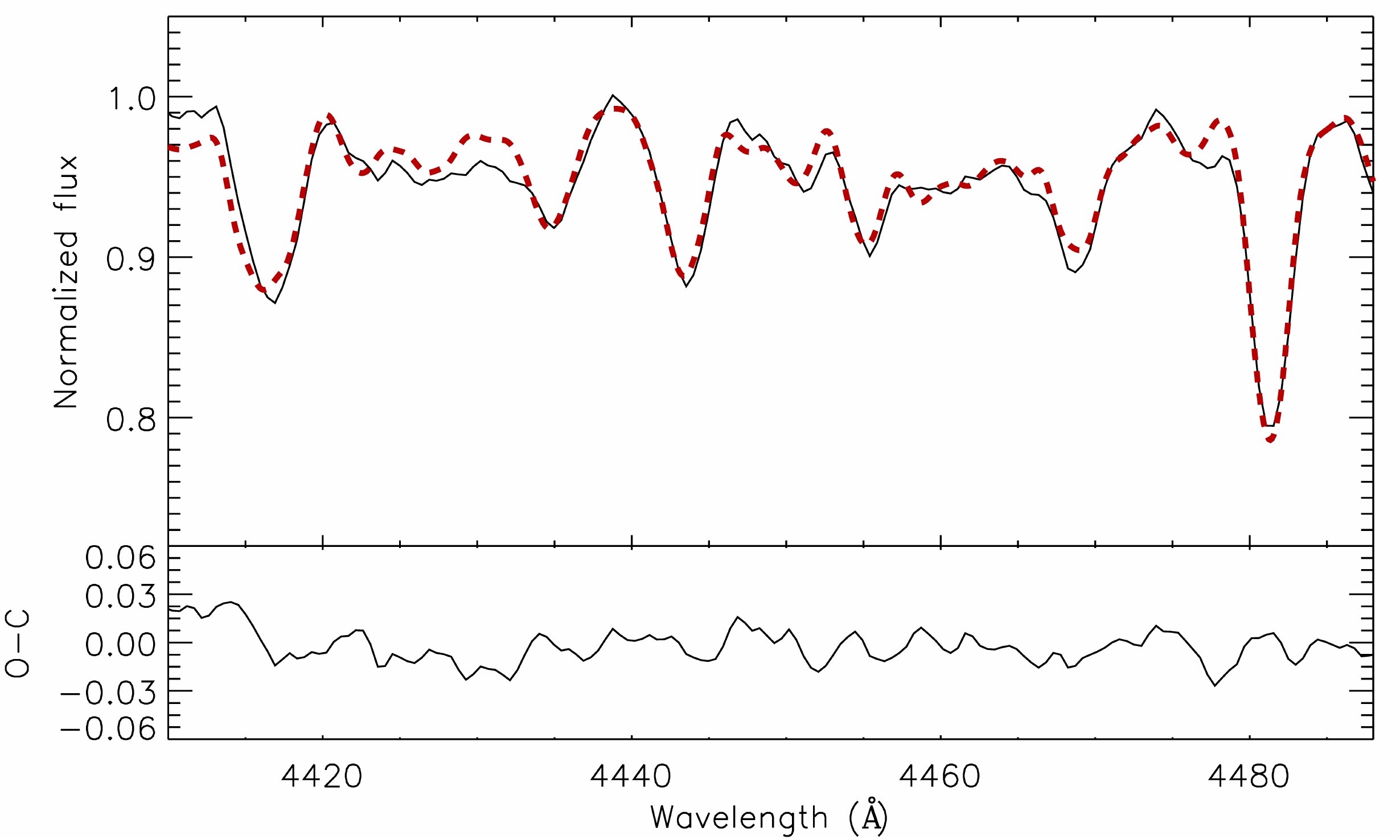}	
\includegraphics[width=1.0\columnwidth]{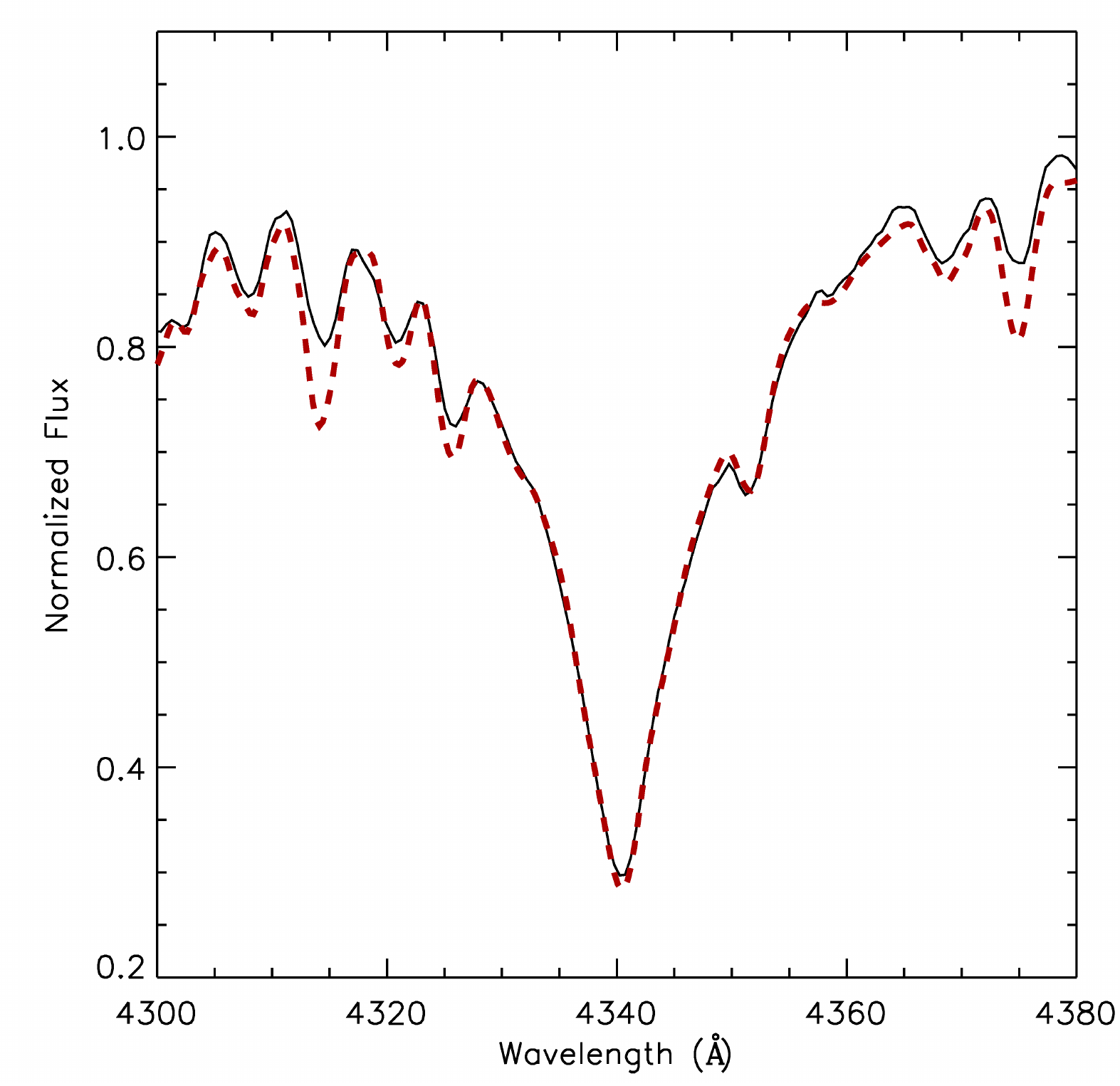}
\caption{Comparison of the synthetic (dashed line) and observed (solid line) spectrum. The best fits to the $H_{\gamma}$ and other lines are shown in the lower and upper panels, respectively.}
\label{fig:spec}
\end{figure}

By then fixing the derived $T_{\rm eff}$ and also the microturbulence value to be 2 km\,s$^{-1}$, we also determined the $\log g$, Fe/H and the $v \sin i$ parameters of the system by applying the same method in the spectral range of $440-460$\,nm. The resulting parameters are given in Table 6. The uncertainties of the parameters were calculated by the  procedure used by \cite{2020MNRAS.493.4518K}. The best fit to the observed spectrum is illustrated in the upper panel of Fig.\,\ref{fig:spec}.

These spectrally inferred stellar parameters are summarized in Table \ref{tbl:spec_parms}.
 
\begin{table}
\centering
\caption{Spectrally Determined Stellar Properties of the Primary}
\begin{tabular}{lc}
\hline
\hline
Parameter & Value   \\
\hline
$T_{\rm eff}$ [K] & $7900 \pm 150$  \\  
$\log g$ [cgs] &  $3.9 \pm 0.2$  \\ 
$[{\rm Fe/H}]/[{\rm Fe/H}]_\odot$ [dex] & $-0.31 \pm 0.14$ \\
$v \, \sin i$  [km/sec] & $113 \pm 6$  \\
\hline
\label{tbl:spec_parms}  
\end{tabular}

\end{table}

\subsection{System Parameters From RV Data Plus SED Fitting}
\label{sec:mcmc}

In order to evaluate the binary system parameters of TIC 63328020, we utilised two essentially independent approaches to the analysis (see also \citealt{2020MNRAS.494.5118K}).  In the first, we find the stellar masses, inclination and system age that best yield a match to the existing measurements of the spectral energy distribution (SED) and the measured radial velocity of the primary star.  In the second approach, we model the {\em TESS} light curve and simultaneously the radial velocity curve with the {\tt phoebe2} binary light curve emulator \citep{2016ApJS..227...29P}.  Both methods utilise a Markov chain Monte Carlo (MCMC) approach to evaluate the uncertainties in the parameters.  

\subsubsection{Coeval, No-Mass-Loss Assumption}
\label{sec:mcmc_coeval}

The first method for finding the system parameters utilises three basic ingredients: (1) the known $K$ velocity for the primary star (see Fig.~\ref{fig:RVs}); (2) the measured SED points\footnote{http://viz-beta.u-strasbg.fr/vizier/sed/doc/; see also Table \ref{tbl:mags}.} between 0.4 and 22 $\mu$m; and (3) the Gaia distance \citep{2018A&A...616A...2L}.  

In this part of the analysis we also make use of the {\tt MIST} ({\tt MESA} Isochrones \& Stellar Tracks; \citealt{2016ApJS..222....8D};  \citealt{2016ApJ...823..102C};  \citealt{2011ApJS..192....3P}; \citealt{2015ApJS..220...15P};  \citealt{2019ApJS..243...10P}) evolution tracks for stellar masses between 0.7 and 3.0\,M$_\odot$ with solar composition\footnote{We have chosen solar metallicity for this part of the analysis due to (i) the uncertainty in the interior vs surface composition of the primary star in TIC~63328020, (ii) the weak spectroscopic determination of $[Z/H] = -0.3 \pm 0.14$ (Table \ref{tbl:spec_parms}), and (iii) the fact that $[Z/H] =0$ turns out to yield the largest number of acceptable models within the wide range of plausible $Z$ values that we explored (see Table \ref{tbl:parms}).  Solar composition for the {\tt MIST} tracks we used was defined by \citet{2016ApJ...823..102C} as: $X_\odot = 0.7154$, $Y_\odot = 0.2703$, and $Z_\odot = 0.0142$ (taken from \citealt{asplund09}).}, in steps of 0.1\,M$_\odot$.  Both here and in Section~\ref{sec:mcmc_no_coeval}, we utilise the \citet{2003IAUS..210P.A20C} model stellar atmospheres for $4000 < T_{\rm eff} < 10,000$\,K in steps of 250\,K.  A solar chemical composition is assumed. 

Our approach follows that of Kurtz et al.~(2020; and references therein), but we briefly describe our procedure here for completeness.  We use an MCMC code (see, e.g.,\,\citealt{2005AJ....129.1706F}) that evaluates four parameters: the primary mass, $M_1$, secondary mass, $M_2$, system inclination angle, $i$, and the {\tt MIST} equivalent evolutionary phase (EEP) of the primary star.  The use of EEPs as a fitted parameter are described in detail in Kurtz et al.~(2020). 

For each step in the MCMC analysis we use the value of $M_1$ and the EEP value for the primary to find $R_1$ and $T_{\rm eff,1}$ from the corresponding {\tt MIST} tracks, using interpolation for masses between those that are tabulated. That also automatically provides an age, $\tau$, for the star.  Since, in this first step of the analysis, we assume that the two stars in the binary are coeval and have experienced no mass exchange, we use the value of $\tau$ to find the EEP for the secondary.  In turn, that yields the values of $R_2$ and $T_{\rm eff,2}$. 

We check to see that neither star overfills its Roche lobe, and if one does, then that step in the MCMC chain is rejected.

The two masses and the orbital inclination angle determine what the $K$ velocity of the primary should be.  This is then compared to the measured value of $85 \pm 5$ km s$^{-1}$, and determines the contribution to $\chi^2$ due to the RV evaluation.

Finally, we use $R_1$ and $T_{\rm eff,1}$, as well as $R_2$ and $T_{\rm eff,2}$, along with interpolated \citet{2003IAUS..210P.A20C} model spectra, to fit the 26 available SED points.  Here $\log g$ is simply fixed at 4.0.  The value of $\chi^2$ for this part of the analysis is added to the contribution from the RV match, and a decision is made in the usual way via the Metropolis-Hastings jump condition (\citealt{1953JChPh..21.1087M}; \citealt{hastings}) as to whether to accept the new step or not.  

This is done $10^7$ times and the posterior system parameters are collected. The parameter posterior distributions are further weighted according to the derivative of the age with respect to the primary EEP number: $d\tau/d({\rm EEP})$ as described in Kurtz et al.~(2020). This corrects for the unevenly spaced EEP points within a larger evolutionary category, and across their boundaries. 

\begin{figure}
\centering
\includegraphics[width=1.0\columnwidth]{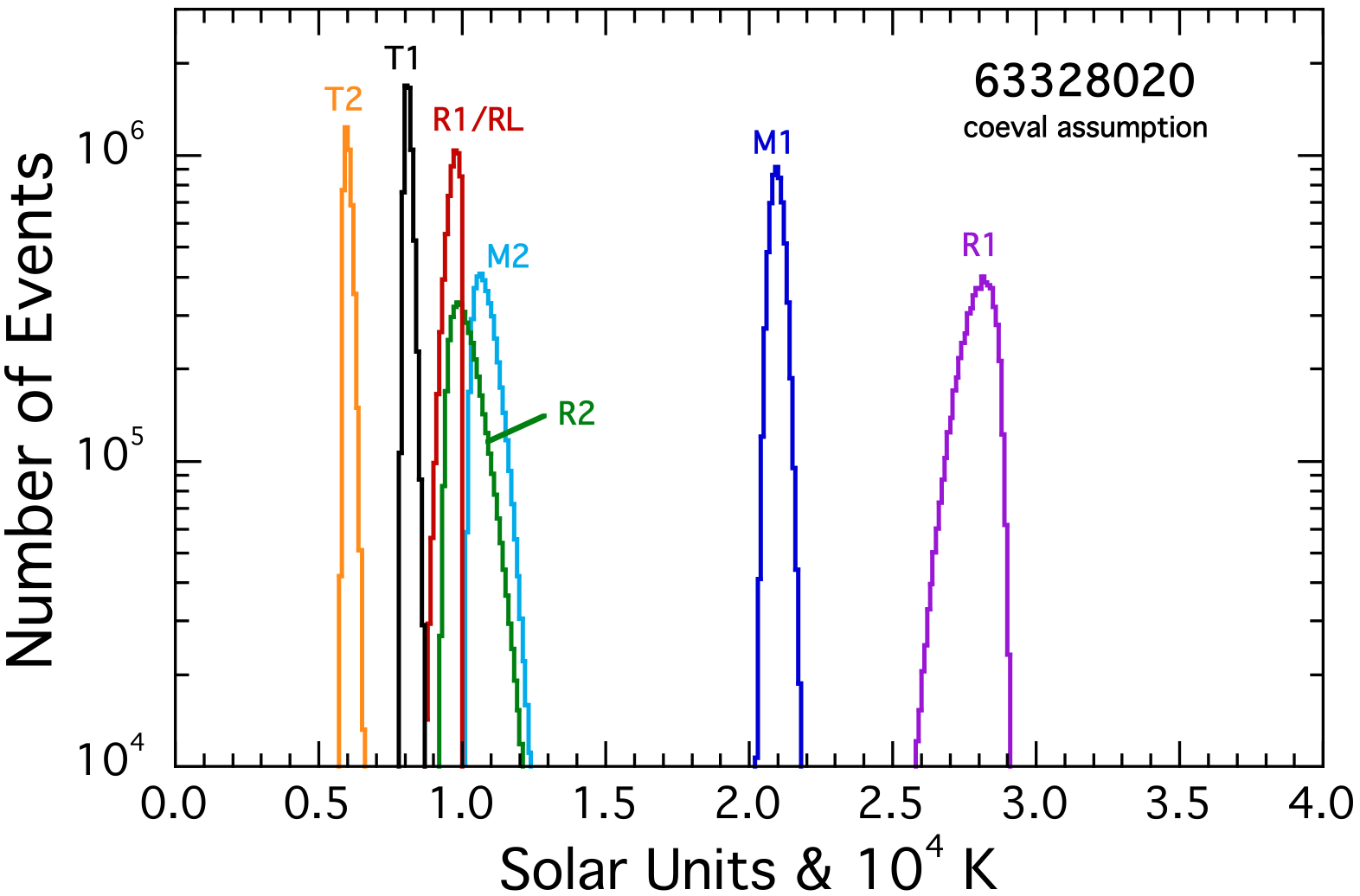}	
\caption{Posterior distributions from the MCMC analysis for the case where the stars are assumed to be coeval and without any prior mass exchange.  For $M_1$, $M_2$, $R_1$, and $R_2$, the x axis is labeled in solar units, while $T_1$ and $T_2$ are in units of $10^4$ K, and $R_1/R_L$ is dimensionless.  Note, in particular, that $R_2$ is small compared to $R_1$.  }
\label{fig:dist1}
\end{figure}

The results of this analysis are summarised in a single plot of distributions in Fig.~\ref{fig:dist1}.  We show the posterior distributions for $M_1$, $M_2$, $R_1$ and $R_2$, in solar units, while $T_1$ and $T_2$ are in units of $10^4$ K, and $R_1/R_L$ is dimensionless ($R_L$ is the radius of the primary's Roche lobe).  For this coeval and no mass exchange scenario, the radius of the lower mass secondary, $R_2$ is much smaller than for the primary, $R_1$.  This results from the fact that if the more massive primary is only somewhat evolved off the ZAMS, then the secondary with a much lower mass cannot be hardly evolved at all.

These results are summarised in the second column of Table \ref{tbl:parms}.

\begin{figure}
\centering
\includegraphics[width=1.0\columnwidth]{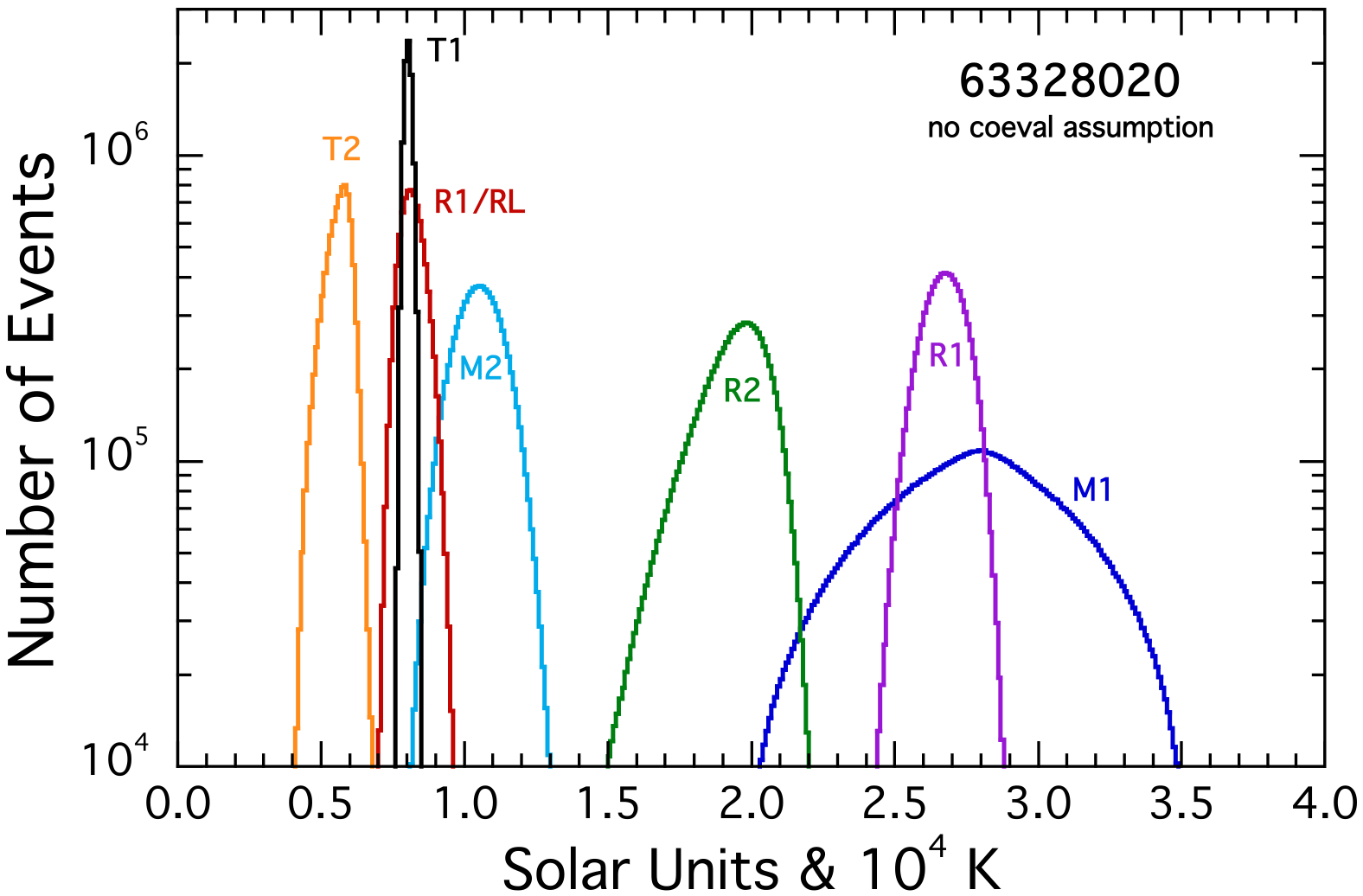}	
\caption{Posterior distributions from the MCMC analysis for the case where the stars are {\em not} assumed to be coeval and without any prior mass exchange.  The axis labeling is the same as in Fig.~\ref{fig:dist2}.  Note, that the distributions are broader than in Fig.~\ref{fig:dist1}, and $M_1$ has gotten larger, as has $R_2$. }
\label{fig:dist2}
\end{figure}  

\subsubsection{Relaxing the No-Mass-Loss Assumption}
\label{sec:mcmc_no_coeval}

Here we utilised the same information as in Section~\ref{sec:mcmc_coeval} (viz, the $K$ velocity of the primary and the SED points, but we relax the constraint that the two stars must be coeval and have undergone no mass exchange.  

The MCMC system parameter evaluation results for this case are summarised by the distributions in Fig.~\ref{fig:dist2}.  We find three major differences from this removal of the coeval constraint: (1) the distributions are considerably broader than in Fig.~\ref{fig:dist1}; (2) the mass of the primary star has shifted considerably to higher values; and (3) the radius of secondary has nearly doubled.

The system parameter results for the case where the no-prior-mass-exchange assumption has been relaxed are summarised in the third column of Table \ref{tbl:parms}.

\begin{table*}
\centering
\caption{Derived Parameters for the TIC~63328020 System}
\begin{tabular}{lccc}
\hline
\hline
Input Constraints & SED + RV$^a$ & SED + RV$^b$ &  Light curve + RV$^c$\\
\hline
Period (days) & 1.1057 & 1.1057 & 1.1057 \\
$K_1$ (km~s$^{-1}$)$^d$ & $85 \pm 5$ & $85 \pm 5$ & $85 \pm 5$ \\  
$v\,\sin i$ (km s$^{-1}$)$^e$ & $113 \pm 6$ & $113 \pm 6$ & ...\\
Spectral & 26 SED points$^f$ & 26 SED points$^f$ & ... \\
Stellar evolution tracks & {\tt MIST}$^g$ & ... & ... \\
Light curve modeling & ... & ... & {\em TESS}$^h$ \\
Distance (pc)$^i$ & $1054 \pm 20$ & $1054 \pm 20$ & $1054 \pm 20$ \\
$A_V$ & 1.0 & 1.0 & 1.0 \\
\hline
Derived  Parameter & SED + RV$^a$  & SED + RV$^b$  & Light curve + RV$^c$\\
\hline
$M_1$ (M$_\odot$) &  $2.10 \pm 0.03$ &  $2.78 \pm 0.35$ & $2.34 \pm 0.10$ \\
$M_2$ (M$_\odot$) & $1.08 \pm 0.04$ & $1.05 \pm 0.10$ & $0.97 \pm 0.05$ \\
$R_1$ (R$_\odot$) & $2.80 \pm 0.06$ & $2.67 \pm 0.09$ &$3.03 \pm 0.05$\\
$R_2$ (R$_\odot$) & $1.01 \pm 0.05$ & $1.95 \pm 0.13$ & $2.03 \pm 0.03$ \\
$T_{\rm eff,1}$ (K) & $8120 \pm 135$ & $8040 \pm 150$ & $8300 \pm 400$ \\
$T_{\rm eff,2}$ (K) & $6000 \pm 140$ & $5660 \pm 460$ & $5650 \pm 250$ \\
$i$ (deg) & $73 \pm 6$ & $76 \pm 4$  & $79.0 \pm 0.4$ \\
$a$ (R$_\odot$) & $6.6 \pm 0.1$ & $7.0 \pm 0.4$ & $6.7 \pm 0.1$ \\
$R_1/R_L$ & $0.97 \pm 0.02$ & $0.82 \pm 0.05$ & $\gtrsim 0.96$ \\
$K_2$ (km~s$^{-1}$)$^j$ & $189 \pm 8$ & $225 \pm 17$ & $204 \pm 6$ \\
age (Myr) & $800 \pm 25$ & ... & ... \\
$\beta_1$$^k$ & ... & ... & $0.73 \pm 0.05$  \\
$A_2$$^l$ & ... & ... & $0.67 \pm 0.05$  \\
\hline

\label{tbl:parms} 
\end{tabular}   

{\bf Notes.}  (a) MCMC fits to the measured RV amplitude plus the SED points.  The assumption is made that the two stars are coeval in their evolution, and have not exchanged any mass. We give the same weight to the $K_1$ `data point' as to any one SED point.  We have also tested other weightings (e.g., weighting the one $K_1$ value several times higher) and it does not change the results significantly. If we had allowed the stellar metallicity of the primary to vary freely instead of fixing it at solar, then we could have at best constrained $Z_\odot/3 \lesssim Z \lesssim 3Z_\odot$.  The corresponding uncertainties in $M_1$, $M_2$, $R_2$, and system age would have increased to $\pm 0.2\, {\rm M}_\odot$, $\pm 0.15\, {\rm M}_\odot$, $\pm 0.15\, {\rm R}_\odot$ and $\pm 300$ Myr, respectively, while $R_1$ and $T_{\rm eff, 1}$ would remain unchanged. (b)  Same as (a) except that the assumption of no prior mass exchange has been dropped. (c) {\tt phoebe2} fit to the {\em TESS} orbital light curve plus the RV amplitude. (d) This work (see Sect.~\ref{sec:phoebe}). (e) Determined from the observed spectra (see Table \ref{tbl:spec_parms}). (f) See Fig.~\ref{fig:sed}. (g) \citet{2016ApJS..222....8D} and \citet{2016ApJ...823..102C}. (h) Modelled with {\tt phoebe2}. (i) Gaia DR2 \citep{2018A&A...616A...2L}.  (j) Predicted from the MCMC parameter evaluations. (k) Gravity brightening exponent.  (l) Bond bolometric albedo of the secondary.

\end{table*}

\subsection{System Parameters From RV Data Plus Light curve Modelling}
\label{sec:phoebe}

In Section~\ref{sec:mcmc_coeval} above we analysed the basic system parameters from an MCMC evaluation of the two masses, the inclination angle, and the evolutionary phases (EEP) of the two stars.  The fitted quantities were $K_1$ and 26 SED points, coupled with the Gaia distance.  In Section~\ref{sec:mcmc_no_coeval} we relaxed the coeval constraint on the two stars and fit independently for their masses and radii.

We now proceed to analyse the system parameters via simultaneous fitting of the {\em TESS} orbital light curve as well as the radial velocity curve using the next-generation Wilson-Devinney code {\tt phoebe2}  \citep{2016ApJS..227...29P,2018ApJS..237...26H,2020ApJS..247...63J,2020ApJS..250...34C}.  First, we removed the pulsations from the light curve.  In addition, a visual inspection of the orbital light curve in Fig.~\ref{fig:lc1} shows that the ELV peak following the primary eclipses is lower than the preceding ELV peak.  This difference can be empirically removed by subtracting a simple sinusoid at the orbital frequency and of amplitude 4110 ppm.  The phasing of this sinusoid would be correct for the Doppler boosting (DB) effect (\citealt{2003ApJ...588L.117L}; \citealt{2010ApJ...715...51V}) if the lower luminosity secondary were the source of the DB.  From the orbital solutions already in hand (see columns 2 and 3 of Table \ref{tbl:parms}), the expected DB amplitude would be $\lesssim700$ ppm for the primary and $\lesssim 300$ ppm for the secondary, and with opposite signs\footnote{The Doppler boosting amplitude for this system should be  $A_{\rm DB} = (\alpha_1 K_1  L_1 - \alpha_2 K_2  L_2) /c L_{\rm tot}$ (\citealt{2010ApJ...715...51V}), where the $L$'s are the luminosities and $\alpha$'s are the Doppler boosting coefficients in the {\em TESS} band.  If we take $\alpha_1 \simeq 2.7 \pm 0.2$, and $\alpha_2 \simeq 4.2 \pm 0.7$, and $L_1/L_{\rm tot}$ = 0.9, then we find $A_{\rm DB} \simeq 400$ ppm.}.  Thus, the observed amplitude is far too large to be the DB effect (which should be only $\sim$400 ppm net), and we attribute it to spots on the secondary that are corotating with the orbit.  Therefore, we elected to subtract off a sinusoidal component with amplitude 4110 ppm from the light curve before carrying out the fitting with {\tt phoebe2}.

The light-curve fitting procedure utilized the MCMC methodology outlined in \citet{2018A&A...619A..84B} and \citet{2019MNRAS.482L..75J}.  The component masses, radii and temperatures, and the orbital inclination were allowed to vary freely over ranges consistent with the observed SED. The only additional free parameters were the gravity brightening exponent, $\beta_1$ (where $T^4_\mathrm{eff,local}=T^4_\mathrm{eff,pole} (g_\mathrm{local}/g_\mathrm{pole})^\beta$), of the primary and the (Bond) bolometric albedo \citep{2019ApJS..240...36H}, $A_2$, of the secondary, which are critical for constraining the ELV and irradiation effect amplitudes, respectively.

\begin{figure}
\centering
\includegraphics[width=1.0\linewidth,angle=0]{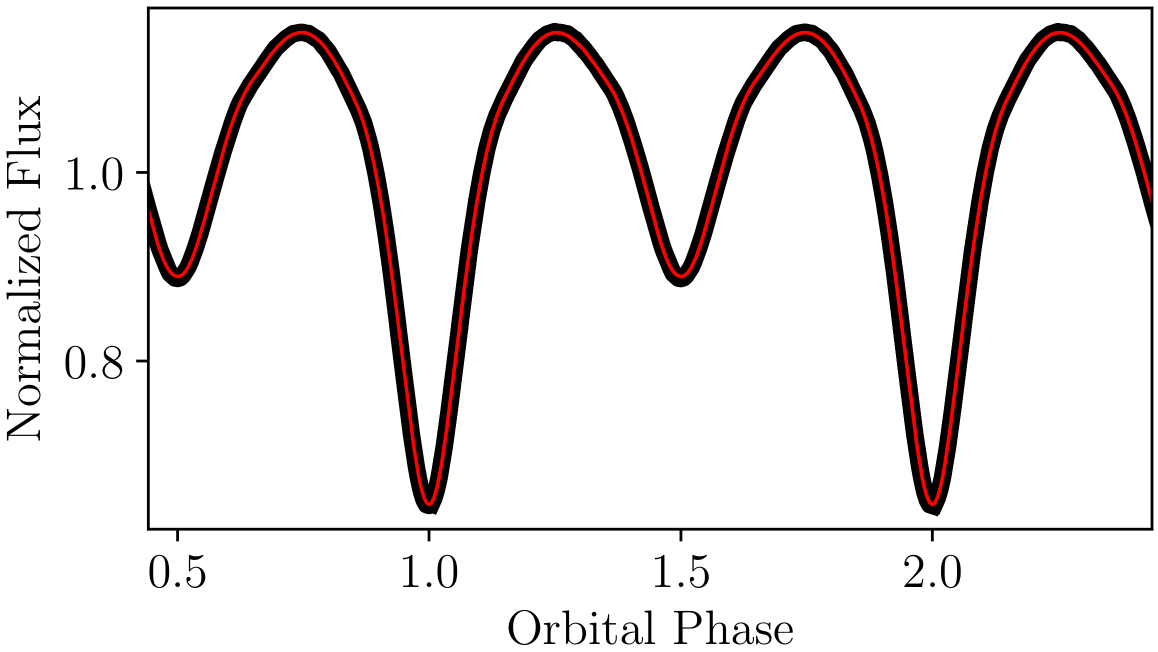}
\caption{The model {\tt phoebe2} light curve (red) on the observations (black). A sinusoidal component of amplitude 0.4 per cent was removed from the {\em TESS} light curve to equalize the two ELV maxima before performing the fit (see text for details).}
\label{fig:phoebe_fit}
\end{figure}

The best-fitting {\tt phoebe2} orbital light curve is shown in Fig.\,\ref{fig:phoebe_fit} while the corresponding model fit to the radial velocities was presented in Fig.~\ref{fig:RVs}.  The resultant model parameters for the system are listed in the last column of Table \ref{tbl:parms}.  

It is clear that the {\tt phoebe2} model provides a remarkably good fit to both the observed light and radial velocity curves, with all model variables extremely well constrained.  The model variables all present with strongly Gaussian posteriors, however several are strongly correlated.  For example, due to the use of a single photometric band, the posteriors of the primary and secondary temperatures show a weak positive correlation.  Likewise, the primary's radius is positively correlated with its mass, with larger primary masses necessitating larger primary radii in order to maintain the same Roche lobe filling factor and thus the same amplitude of ELV.  Ultimately, further observations are required to break these correlations but, nonetheless, the current data are sufficient to strongly constrain the properties of the system \citep[see][for further discussion of the fitting of a single-band ELV light curve for a similar case, albeit without the observed eclipses of TIC~63328020 which provide additional strong constraints]{2020MNRAS.494.5118K}.

The results for the system parameters derived from the {\tt Phoebe} fit to the {\em TESS} are summarised in the fourth column of Table \ref{tbl:parms}.

\section{Formation and Evolution -- Prior History of Mass Transfer}
\label{sec:mdot}

In order to determine the formation history of TIC~63328020, we must first consider whether or not mass transfer between the binary components has had a significant effect on its evolution. The second column of Table \ref{tbl:parms} lists the inferred properties of TIC~63328020 that were derived based on the SED and RV data under the assumption of no mass transfer during the binary's evolution. With this latter constraint relaxed, the properties of the binary deduced using (i) the SED and RV data, and (ii) the RV data in conjunction with a {\tt phoebe2} fit to the light curve, are shown in the third and fourth columns, of Table \ref{tbl:parms}, respectively. 

The most glaring discrepancy between any of the predicted properties of the two stars occurs for the radius of the secondary ($R_2$). The inference for the radius made under the assumption of no mass transfer disagrees with the other two inferences by nearly a factor of two. Given that the last two inferences were derived independently (although they do share the same RV data) and given that no constraint on mass transfer was imposed, the relatively good agreement between these two cases seems to imply that TIC~63328020 very likely experienced mass transfer in the past. Moreover, an orbital period on the order of days is typical of many `Algol-like' binaries for which mass transfer/loss occurred during their prior evolution (see, e.g., \citealt{1989SSRv...50.....B}; \citealt{2000NewAR..44..111E}).

In analysing the evolution of the progenitor binary we will therefore assume that mass transfer occurred. The next question to address is whether or not the current primary was the original primary of the progenitor system (i.e., the more massive one) or whether a mass-ratio reversal occurred (i.e., an Algol-like evolution). If a mass-ratio reversal did not occur, then we are forced to conclude that the original primary could not have lost much mass simply because the original mass ratio ($M_1/M_2$) would have been so large ($\gtrsim 3$) that the binary would have undergone a dynamical instability leading to the presumed merger of the two stars (see, e.g., \citealt {1976ApJ...209..829W}; \citealt{1997A&A...327..620S} for a discussion of the conditions leading to dynamical instability). 

For the other scenario, the masses of the two primordial components (we will refer to them as `Star 1' and `Star 2') may have changed significantly during the course of the evolution. The largest uncertainty concerns the degree to which mass transfer between the components is conservative. For a fully conservative transfer, all of the mass that is lost by the more massive star (Star 1) is subsequently accreted by Star 2. On the other hand, for completely non-conservative mass transfer, the mass of Star 2 would not change as the mass of Star 1 decreased. Because we do not know how non-conservative mass transfer could have been (or the amount of angular momentum transported out of the binary), we have investigated a realistic range of possibilities.

\begin{figure*}
\centering
\includegraphics[width=0.80\textwidth,angle=0]{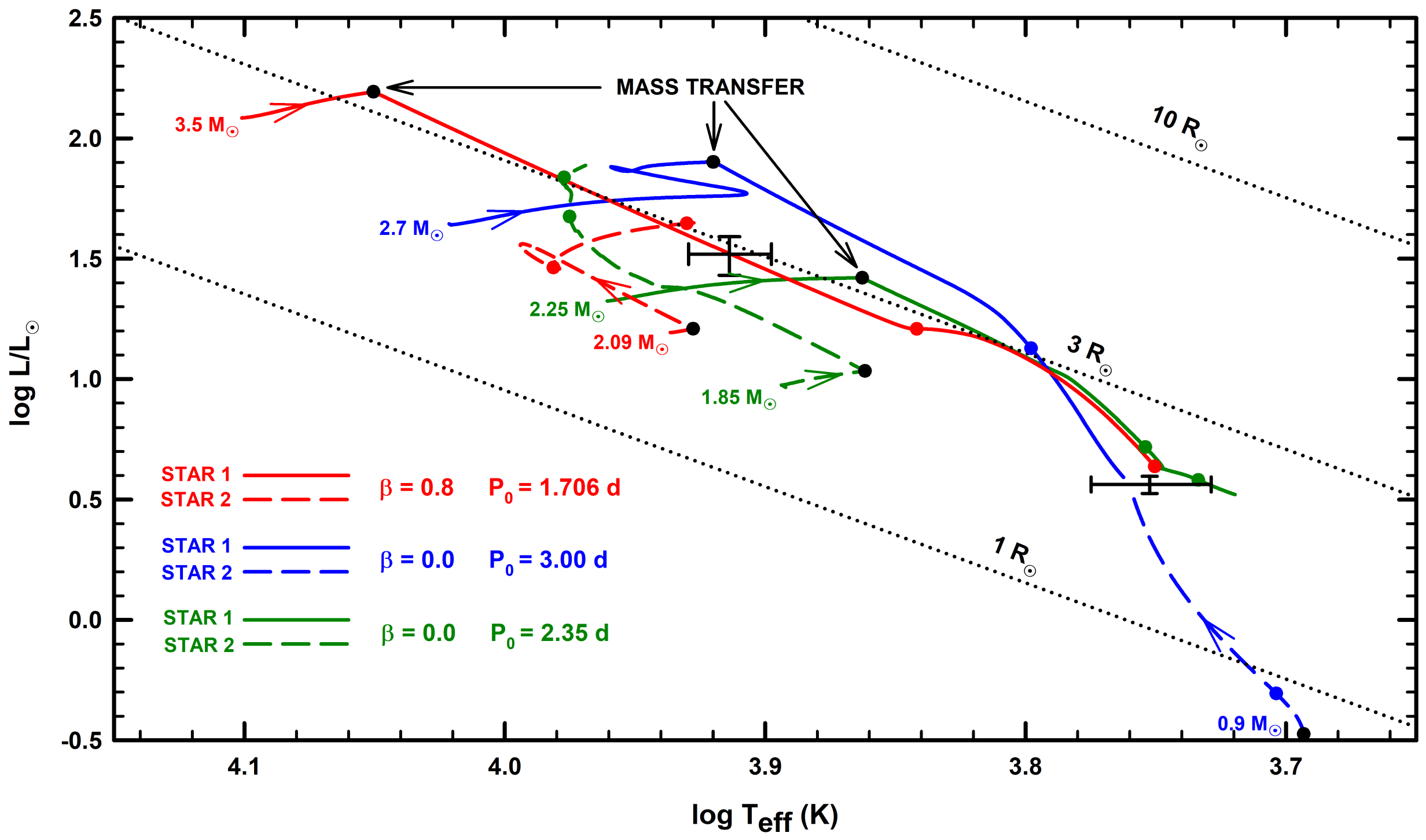}
\caption{Representative tracks of three distinctly different primordial binary evolutions leading to the formation of TIC~63328020 are shown in the HR~diagram. The evolution of the more massive component of the primordial binary (Star 1) is denoted by a solid curve and the evolution of its less massive companion (Star 2) is represented by a dashed curve. Arrows superposed on the curves denote the direction of increasing age (i.e., the direction in which the component is evolving). Different colours are used to denote each pair of the components of the primordial binary. Each component initially starts evolving from the ZAMS and those masses for all three cases are labelled. The red tracks correspond to a binary that experiences highly non-conservative mass-transfer ($\beta = 0.8$), while the blue and green tracks assume fully conservative mass-transfer.  The solid dots of the same colour as the track indicate the points in the evolution when the orbital period matches the observed value (1.1057\,d). The black dots indicate the points for which (rapid) thermal-timescale mass-transfer is first initiated. All of the tracks terminate when both stars have simultaneously fill their Roche lobes. The error bars mark the current observed luminosity and $T_{\rm eff}$ for both components of the system. }
\label{fig:evol1}
\end{figure*}  

\subsection{Evolutionary Grid}
\label{sec:gridsl}

To determine the properties of putative primordial binaries that could evolve to approximately match the observational properties inferred for TIC~63328020,  we have computed evolutionary tracks for an extremely wide range of initial conditions and assumed input physics.  To optimise the numerical computations, we were guided by a number of grids that had been previously generated to solve for the evolution of other types of interacting binaries. Specifically, we have used the grids generated for post-Algol binaries such as MWC882 \citep{2018ApJ...854..109Z} and wide, hot subdwarf binaries \citep{2019AAS...23421502N} to try to constrain the range of possible initial conditions.  Once this was accomplished, additional (more precise) grids were successively computed until we were able to enumerate a reasonably precise set of primordial binaries that could evolve to produce reasonable facsimiles of TIC~63328020.

The evolutionary tracks were calculated using the binary version of {\tt MESA}{\footnote {The results presented in this paper were computed with release 10108.}} for which the evolution of both the donor and accretor stars are computed simultaneously (see \citealt{2011ApJS..192....3P}; \citealt{2015ApJS..220...15P}; \citealt{2019ApJS..243...10P}).  The grids cover a range of initial conditions describing the properties of the primordial binaries.  Specifically, we created grids for primary masses (i.e., Star 1, the more massive component) in the range of  $1 \le M_{1,0}/{\rm M}_\odot \le 4$.  The mass of the secondary (Star 2) was expressed in terms of the mass ratio ($q$) of the primary's mass to the secondary's  mass.  We explored the range of $1.05 \le q_0 \le 4$. Finally, the primordial orbital period was expressed in terms of the critical period ($P_{\rm c}$)  for which the primordial primary would just be on the verge of overflowing its Roche lobe.  Orbital periods in the range of $1 \le P_{\rm {orb}}/P_{\rm c} \le 10$ were computed.  In all, some 1800 new binary evolution models were generated, in addition to the original $\simeq 4000$ that we already had in our library of computations for Algol-like systems.

We also investigated the effects of metallicity.  Given the binary's proximity to the mid-plane of the Galaxy, we chose values in the range of $0.01 \le Z \le 0.03$ which is a reasonable range for Population I stars.  We found that a metallicity of $Z=0.03$\footnote{This value of $Z$ does not have to be same as used to compute the 2nd column in Table \ref{tbl:parms} since those results, which assume a coeval evolution, turn out to be invalid, regardless of $Z$.} best matched the observations of the effective temperatures, and for this reason we adopted that value when computing the final grid of models (with $X = 0.693$ and $Y = 0.277$)\footnote{These {\tt MESA} values are in excellent agreement with those of \citet{coelho07} as interpolated from their Table 1: $X = 0.689$, $Y = 0.281$, $Z=0.030$.}.   We will return to this issue later and discuss our choice for the metallicity.

In terms of the input physics, the degree to which mass transfer is non-conservative and the mechanism describing the systemic loss of orbital angular momentum is very uncertain. This uncertainty can be parametrized in terms of the quantities $\alpha$ and $\beta$ (for details see \citealt{2006csxs.book..623T}).  In the {\tt MESA} code, $\alpha$ is the fraction of the mass lost by the donor star that gets ejected from the binary such that the mass carries away the specific angular momentum of the donor star (i.e., fast Jeans' ejection), and $\beta$ is the fraction of the mass that is ejected from the accretor and is assumed to carry away the specific angular momentum corresponding to that star.  Thus the mass gained by the accretor (secondary) can be written as:
\begin{equation}
\delta M_{2}=-(1-\alpha -\beta)\delta M_{1}.
\label{321}
\end{equation}
We further assume that none of the mass that is lost from the binary forms a circumbinary torus that can extract additional orbital angular momentum during the binary's evolution. 

Given the uncertainty in the values of $\alpha$ and $\beta$, our evolutionary tracks were computed for a range of values such that $0 \le \alpha \le 0.6$ and $0 \le \beta \le 1$, under the constraint that $\alpha +\beta \le 1$.  We draw the qualitative conclusion that the sum of $\alpha +\beta$ has a much greater effect on the evolutionary outcomes than the combination of individual choices of $\alpha$ and $\beta$ that give the same sum.\footnote{Note that the choices of the primordial component masses and orbital period will have a profound effect on the evolution.} Thus to minimise numerical computations, our final set of models has been computed with $\alpha = 0$.  Finally, orbital angular momentum dissipation was calculated based on the torques associated with gravitational radiation and magnetic braking as described in \citet{2015ApJ...809...80G} and \citet{2016ApJ...833...83K}, with the magnetic braking index set equal to 3.  The magnetic braking formula (Verbunt-Zwaan law; \citealt{verbunt81}) was inferred from observations of low-mass main-sequence stars and thus must sometimes be extrapolated to stars that are either evolved, very low-mass, or rapidly rotating. Although the magnitude of magnetic braking torques remains uncertain, it has relatively little effect on the evolutionary tracks until after thermal timescale mass transfer has occurred.  

After generating our grid of binary evolution tracks, we found that both conservative and non-conservative evolutions could produce the desired results given the appropriate choices of the primordial masses and the primordial period. Thus we conclude that a fine-tuning of the initial conditions is not required in order to reproduce the observations.  Possible evolutionary scenarios can be divided into two separate classes: (1) the more massive star (Star 1) loses a relatively small fraction of its initial mass while the companion (Star 2) gains some portion of that mass; or, (2) the more massive primordial star loses a large fraction of its mass leading to a mass-ratio reversal (the mass ratio being defined as $q = M_{\rm Star1}/M_{\rm Star2}$), thus implying that the accretor becomes more massive than the donor. For either scenario, the evolution can be fully conservative ($\beta = 0$) or highly non-conservative ($\beta = 0.8$).

We find that the first scenario (mass ratio does not change significantly) never fully reproduces the observationally inferred properties listed in columns 2 and 3 of Table \ref{tbl:parms}. Although this grouping (class) of evolutionary tracks can reproduce most of the properties of TIC~63328020, we did not find a primordial binary that could ultimately produce a secondary star with such a large radius ($\approx 2 {\rm R}_{\odot}$) while simultaneously matching all of the other inferred properties of both stars. The problem stems from the following physical constraints: (i) in order to bloat the accretor sufficiently, mass-transfer rates in excess of $\sim 3 \times 10^{-7} \,{\rm M}_\odot \,{\rm yr}^{-1}$ are required for extended periods of time; and, (ii) binaries with large mass ratios tend to experience dynamical instabilities when Roche lobe overflow first commences.  With respect to the latter issue, the very large mass ratio inferred for TIC~63328020 necessarily implies that the accretor could only have gained a few tenths of a solar mass during the evolution (otherwise the initial evolution would have been dynamically unstable). And given the required high mass-transfer rates and the small net accretion, this implies that mass transfer would have occurred over an extremely short interval ($\lesssim 1$ Myr), making the whole scenario less likely. Moreover, mass-transfer rates of $\sim 10^{-6} \,{\rm M}_\odot \,{\rm yr}^{-1}$ are expected at the current epoch and there is little evidence to support such a high value (see the discussion below).

According to the second scenario, the original primary of the primordial binary (i.e., the donor) loses so much mass to its accreting companion that a mass-ratio reversal occurs (in other words, the observed low mass secondary of TIC~63328020 was originally the higher mass star). As discussed above, the largest uncertainty concerns the choice of $\beta$ and we attempt to mitigate the effects of this uncertainty by creating a grid of models with the variable $\beta$ taken to be one of the dimensions of parameter space. It is generally expected that the evolution of Algol-like binaries will be at least mildly non-conservative (see, e.g., Eggleton 2000) and that is why we chose to investigate the range $0 \leq \beta \leq 0.8$.

The evolution of both binary stars in the Hertzsprung-Russell diagram for three representative systems is shown in Fig.~\ref{fig:evol1}.  The blue curves illustrate the first scenario, while two sets of tracks represent the second scenario---corresponding to extreme values of $\beta$, i.e., $\beta = 0$ and 0.8, green and red curves, respectively.  For the first scenario (see the solid and dashed blue curves for the evolution of the donor and accretor, respectively), we chose a primordial binary consisting of 2.7 and 0.9\,M$_\odot$ components with an orbital period of 3.0\,d. Possible solutions for TIC~63328020 are denoted by the solid blue dots. For the second scenario, this fully conservative case has components initially consisting of 2.25 and 1.85\,M$_\odot$ stars in a 2.35-d orbit (see the green solid and dashed curves, respectively). The highly non-conservative evolution with $\beta = 0.8$ is denoted by the red curves. 

The primordial binary consisted of a 3.5\,M$_\odot$ donor in a 1.706-d orbit with a 2.09\,M$_\odot$ accretor. For these latter two sets of evolutionary tracks there are two sets of solid dots (green and red) that denote possible solutions at the observed orbital period of 1.1057\,d. In each case, the latter set of dots (corresponding to a later age) better fits the inferred properties of TIC~63328020 enumerated in Table \ref{tbl:parms}. The solid black dots indicate the onset of (rapid) thermal timescale mass transfer. For all cases, the evolutionary tracks are seen to abruptly change their trajectories in the HR diagram once mass transfer commences. The donor stars all tend to evolve towards lower luminosities and effective temperatures while the accretors immediately evolve towards higher temperatures and luminosities.  

Each of the three tracks terminates once the accretor has expanded sufficiently to fill its Roche lobe.  A summary of the three tracks and the best fit to the inferred data enumerated in columns 3 and 4 of Table \ref{tbl:parms} is presented in Table \ref{tbl:model}.  Note that the subscripts 1 and 2 denote the properties of the primary and secondary, respectively, for TIC~63328020 at the current epoch.  The age is measured from the formation of the primordial binary and $\log \dot M$ indicates the (current) mass-transfer rate from the donor star.

\begin{figure}
\centering
 \hglue-0.15cm \includegraphics[width=1.02\linewidth,angle=0]{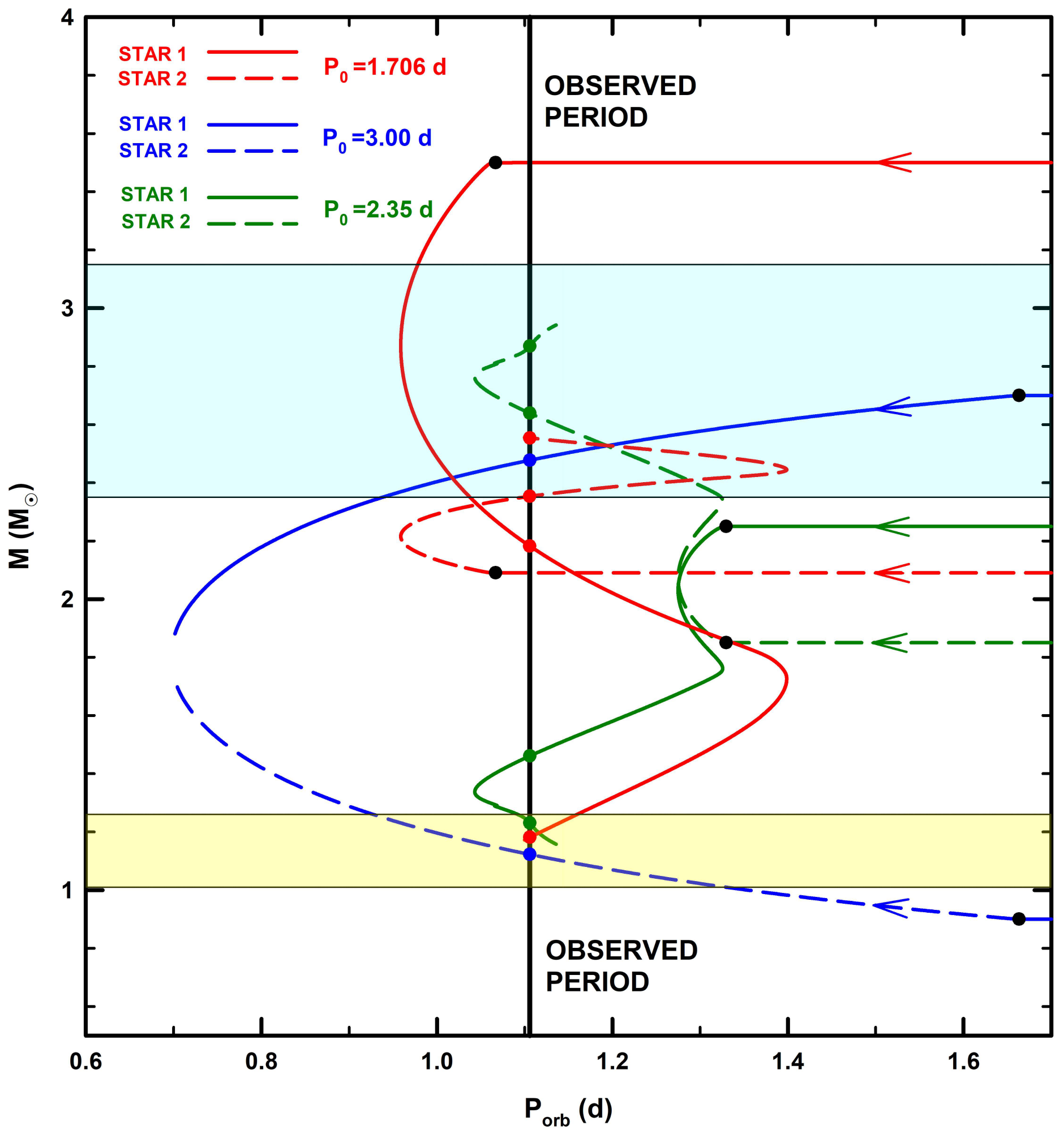}
\caption{The evolution of the masses of the two components of the primordial binary as a function of orbital period is illustrated. Arrows superposed on the curves denote the direction of increasing age (i.e., the direction in which the component is evolving). Star 1 is always losing mass while Star 2 is accreting mass. The colour scheme for the three cases shown in Fig.~\ref{fig:evol1} is repeated here. The initial orbital period for each case is also labeled (upper left) and the observed orbital period (1.1057\,d) is denoted by a vertical black bar. The solid dots of the same colour as the curves indicate the points that could be possible solutions for the properties of TIC~63328020. The cyan-shaded region represents the approximate range of possible masses of the current primary based on the analysis presented in Table \ref{tbl:parms} (columns 3 and 4). The yellow-shaded region corresponds to the possible range of masses for the current secondary. All of the curves terminate when both stars have simultaneously filled their Roche lobes. }
\label{fig:evol2}
\end{figure}  

In order to further elucidate the scenarios associated with our three representative tracks, the evolution of the masses of each component is shown as a function of the orbital period ($P_{\rm orb}$) in Fig.~\ref{fig:evol2}. The colour coding and the use of solid and dashed lines in addition to the solid dots have the same meaning as that described for Fig.~\ref{fig:evol1}.  For track \#1 (blue curves), the primordial donor simply loses a few tenths of a solar mass that is then gained by the primordial secondary (accretor).  Although this is one of the simplest types of evolution that could reproduce the observed masses of TIC~63328020, we were unable to find any combination of initial conditions or values of the parameters governing some of the input physics (e.g., orbital angular momentum dissipation) that reproduced the inferred radius of the secondary.  Instead, both tracks \#2 (green curves) and \#3 (red curves) can produce reasonable facsimiles of TIC~63328020.  

Both donor stars (solid curves) initially undergo thermal timescale mass transfer causing the orbital period to decrease (see Fig.~\ref{fig:evol2}) as both donor stars shrink due to quasi-adiabatic mass loss.  However, once the mass ratio has been reduced to about unity, further mass transfer causes the orbits to expand (with a concomitant increase in $P_{\rm orb}$ as the donor stars readjust thermally.  As the donor stars approach quasi-thermal equilibrium (with a much reduced mass transfer rate), they contract forcing the orbit to shrink{\footnote{Note that the green track of the donor star subsequently experiences a second orbital period minimum.  This is due to the thermal re-adjustment of the core from purely convective energy transport to a fully radiative mode.}.  The accretors for both cases continuously gain mass with a resulting increase in their radii.  The tracks terminate once the accretors also fill their Roche lobes.  

\begin{table}
\centering
\caption{{\tt MESA} Model Parameters for the TIC~63328020 System}
\begin{tabular}{lccc}
\hline
\hline
Model Parameter  &	Track \#1 & Track \#2 & Track \#3 \\
                          &	(Blue)       &  (Green)   & (Red)        \\
\hline
$M_{\rm Star1,0}$ (M$_\odot$)$^a$ & 2.70 & 2.25 & 3.50 \\
$M_{\rm Star2,0}$ (M$_\odot$)$^a$ & 0.90 & 1.85 & 2.09 \\
$P_{\rm orb,0}$ (d)$^a$ & 3.00 & 2.35 & 1.71 \\
$\beta$$^b$ &  0.0 & 0.0 & 0.8 \\
$q_0 (\equiv M_{\rm 1,0}/M_{\rm 2,0})$$^c$ & 3.00 & 1.216 & 1.675 \\
$M_1$ (M$_\odot$)$^d$  & 2.48 & 2.87 & 2.55 \\
$M_2$ (M$_\odot$)$^d$ & 1.12 & 1.23 & 1.18 \\
$R_1$ (R$_\odot$)$^d$ & 3.10 & 3.07 & 3.06 \\
$R_2$ (R$_\odot$)$^d$ & 0.92 & 2.22 & 2.19 \\
$T_{\rm eff,1}$ (K)$^d$ & 6280 & 9490 & 8510 \\
$T_{\rm eff,2}$ (K)$^d$ & 5050 & 5410 & 5630 \\
$L_{\rm bol,1}$ (L$_\odot$)$^d$ & 13.4 & 68.7 & 44.2 \\
$L_{\rm bol,2}$ (L$_\odot$)$^d$ & 0.49 & 3.80 & 4.33 \\
$R_1 /R_{\rm L,1}$$^d$& 1.0 & 0.94 & 0.98 \\
$R_2 /R_{\rm L,2}$$^d$ & 0.43 & 1.0 & 1.0 \\
$\log \dot M$ (M$_\odot \, {\rm yr}^{\rm -1}$)$^e$ & $-$5.57 & $-$8.69 & $-$9.00 \\
$\dot P_{\rm orb}/P_{\rm orb}$ (${\rm yr}^{\rm -1}$) & $-4.0$E-6& $+5.9$E-10 & $-6.1$E-10 \\
system age (Myr) & 485 & 935 & 545 \\
\hline

\label{tbl:model} 
\end{tabular}  

\hfill\parbox[t]{8.4cm} {{\bf Notes}. (a) `Star 1' and `Star 2' refer to the original primordial primary and secondary, respectively. The subscript `0' indicates the initial system parameters. (b) $\beta$ is the fraction of mass that is transferred to the accretor but is ejected from the system with the specific angular momentum of that star.  The parameter $\alpha$ (not in the Table) is fixed at 0.0 and is the fraction of mass lost by the donor star that is ejected from the system with the specific angular momentum of the donor star.  (c) Initial mass ratio of the primordial binary ($M_{\rm Star,1}/M_{\rm Star,2}$).  (d) These are the model parameters for the current-epoch TIC~63328020 system. (e) The total rate of mass lost by the donor star. }

\end{table}

\subsection{Discussion}
\label{sec:discuss}

Based on the analysis of an extensive grid of models, we conclude that there is a wide range of initial conditions that can replicate the currently observed properties of TIC~63328020.  The simplest type of evolution wherein the primordial primary loses a few tenths of a solar mass to a much less massive accretor, although appealing, cannot reproduce all of the inferred properties. But a wide range of evolutionary scenarios for which a mass-ratio reversal occurs (i.e., the primordial primary becomes the less massive secondary) can be accommodated. In particular, if the evolution is highly non-conservative, then the total mass of the primordial binary would have to be considerably more massive than the presently inferred value with an initial mass ratio of $q_0 \gtrsim 1.5$. For more conservative evolutions, the initial total mass can be much smaller and the mass ratio much closer to unity. Thus there is a wide range of initial parameters for the primordial binary that can produce robust models of TIC~63328020.  

Although there are significant uncertainties associated with systemic mass loss and the orbital angular momentum dissipated as a result of this non-conservative mass transfer, we find that the individual choices of the parameters $\alpha$ and $\beta$ are not nearly as important as the contribution from their sum. For this reason, we parametrized the effects of non-conservative mass transfer in terms of $\beta$ ($\alpha = 0 $). We conclude that the values of $\beta$ in the range of 0 to 0.8 can lead to plausible solutions for the properties of TIC~63328020 (see columns 3 and 4 of Table \ref{tbl:parms}). However, the highly conservative models tend to produce primaries with higher effective temperatures ($ \gtrsim 1000$\,K higher). For this reason, we somewhat prefer models for which $\beta \simeq 0.3$.  

Another important result to note is that the `simple' evolutionary scenario implies that mass-transfer rates at the present epoch should be on the order of $10^{-6} \, {\rm M}_\odot \,{\rm yr}^{-1}$.  By way of contrast, the `mass-reversal' scenario requires mass-transfer rates that are typically three orders of magnitude smaller (Table \ref{tbl:model}).  We examined the spectrum of TIC~63328020 for P Cygni profiles and found no evidence for that feature. This would seem to imply a relatively low mass transfer rate. We also examined four WISE band observations looking for any evidence of nebulosity that might be expected due to a significant wind emanating from the binary. We could not find any hint of nebulosity in that region. There was also no sign of any NIR nebulosity from the PanSTARRS images. Although not conclusive, these results seem to hint at a relatively low mass transfer rate or one that is not highly non-conservative.

One of the very intriguing features of many of our evolutionary tracks that reproduce robust models of TIC~63328020 is that the accretor is very close to filling its Roche Lobe.  Since it is relatively unlikely that we would find such a configuration based solely on the observed pulsational properties, the question arises as to whether the binary had already evolved to a point where both stars temporarily over/filled their Roche lobes  before one of them contracted leading to the currently observed configuration.  If one of the stars contracted then it is possible that it could remain in a detached state for at least a Kelvin time. On the other hand, it is quite possible that the binary would have merged were contact to have occurred.  We are currently trying to address this and questions related to the formation and evolution of WU Ma binaries using smoothed particle hydrodynamics (SPH; S.~Tripathi, L.~Nelson, \&  T.\,S.~Tricco 2020 [in preparation]). 

Finally, we comment on our choice of generating the evolution tracks with a metallicity of $Z = 0.03$.  In the process of selecting an appropriate $Z$, we have explored the effects of metallicity on the evolution of representative models describing the observed properties of TIC 63328020.  Specifically, we generated grids of models for the mass fraction of metals in the range of $0.01 < Z < 0.03$ corresponding to between 60 per cent and 170 per cent of the estimated solar value. We conclude that in order to reproduce the type of evolution described by Track \#1 (see Figure \ref{fig:evol1} and Table \ref{tbl:model}), low values of the metallicity ($Z \lesssim 0.02$) produce secondary masses that are too massive by factors of 25 per cent compared to what is expected based on the results presented in Table \ref{tbl:parms}. For Tracks \#2 and \#3, the difficulty is that, as the metallicity is decreased, the effective temperature of the primary becomes unpalatably large. For example, taking $Z=0.02$, we find that $T_{\rm eff}$ increases by $\sim$400 K for Track \#2 and 1200 K for Track \#3 (to about 10,000 K in each case). Based on the analysis presented in Table \ref{tbl:parms} where the effective temperatures are in the range of $\sim$8000 to 8200 K, we believe that the higher metallicity tracks do a better job of reproducing the observationally inferred properties of TIC 63328020.

Moreover, based on the kinematics and location of TIC 63328020 in the Galaxy, we conclude that it is consistent with a relatively young (extreme) Population I metallicity. Using Gaia's estimated distance of 1050 pc and the galactic latitude of 1.18$^\circ$, its distance above the galactic mid-plane is only about 20 pc (the scale height of the thin disk being about 300 pc). Based on Gaia's estimate of the tangential velocity and using our radial velocity of the binary's centre of mass ($\gamma = 8.2$ km s$^{-1}$ in Table 5), we estimate the spatial velocity to be between $\sim$15 and 20 km s$^{-1}$. These properties suggest that TIC 63328020 could well be a young, high metallicity Population I system. It is also worth noting that all of our evolutionary models -- including the low-metallicity ones described above -- suggest a relatively young age of $<$ 1 Gyr (see Table \ref{tbl:model}). Given that TIC 63328020 has the hallmarks of a high-metallicity Population I system, we adopt $Z=0.03$ as a reasonable value for the metallicity.

\section{Conclusions}

In this work we report the discovery of a short-period binary with tidally-tilted pulsations at $\nu = 21.09533$ d$^{-1}$.  The pulsation amplitude varies with orbital phase and is a maximum at orbital quadrature, i.e., when the ellipsoidal light variations are at a maximum.  The phase of the pulsations rapidly change by more than $\pi$ radians around the time of the primary eclipse, and there is a smaller jump in phase at the secondary eclipse by about half that amount in the opposite direction.  We note that the phase is not a pure $\pi$-radian jump because the mode is distorted from a pure sectoral dipole mode.

In order to help visualize how the tidally tilted pulsations would appear to an observer on an circumbinary planet orbiting TIC 63328020, we include a simulation in the form of an MP4 video (`TIC63328020.mp4').  The video is supplied as Supporting Information for the paper. This same video is also presented in \citet{2020MNRAS.tmp.2716F}.

The pulsating star has $M_1 \simeq 2.5 \, {\rm M}_\odot$, $R_1 \simeq 3 \, {\rm R}_\odot$, and $T_{\rm eff,1} \simeq 8000$ K, while the secondary has $M_2 \simeq 1.1 \, {\rm M}_\odot$, $R_2 \simeq 2 \, {\rm R}_\odot$, and $T_{\rm eff,2} \simeq 5600$ K.  Both stars appear to be close to filling their respective Roche lobes. The orbital period is constant to a part in $\sim$$10^5$ over the last century. However, the period appears to vary erratically on timescales of weeks to decades. At present we have no firm explanation for this behavior.  

We have carried out an investigation of the history of this system with an extensive set of binary evolution models.  We conclude that the most likely scenario is that there has been a prior epoch of mass transfer which has reduced the mass of the original primary so that it is currently the low-mass secondary.  By contrast, the original secondary is now the pulsating primary star.  The mass transfer may still be ongoing with a low mass-transfer rate of $\sim$$3 \times 10^{-9} \, M_\odot$ yr$^{-1}$.
 
Although the architecture and evolutionary histories of the three known tidally tilted pulsators (HD 74423, CO Cam, and TIC~63328020) are unique, they all feature tidally distorted $\delta$~Sct pulsators in short-period orbits. Whereas HD 74423 and CO Cam feature axisymmetric tidally tilted pulsations trapped on the L$_1$ side of the star, the tidally tilted mode in TIC~63328020 exhibits very different phase and amplitude modulation, indicative of a non-axisymmetric ($|m|=1$) mode that is not completely trapped on either side of the star \citep{2020MNRAS.tmp.2716F}.  Future discoveries of tidally tilted pulsators will likely reveal more diversity amongst this new class of stars.

\section*{acknowledgements}

We are grateful to an anonymous referee whose comments and suggestions helped clarify the presentations in this paper.

This paper includes data collected by the TESS mission.  Funding for the TESS mission is provided by the NASA Science Mission directorate. Some of the data presented in this paper were obtained from the Mikulski Archive for Space Telescopes (MAST). STScI is operated by the Association of Universities for Research in Astronomy, Inc., under NASA contract NAS5-26555. Support for MAST for non-HST data is provided by the NASA Office of Space Science via grant NNX09AF08G and by other grants and contracts.

Based on observations made with the Isaac Newton Telescope operated by the Isaac Newton Group of Telescopes, which resides on the island of La Palma at the Spanish Observatorio del Roque de los Muchachos of the Instituto de Astrof\'isica de Canarias.  The authors thankfully acknowledge the technical expertise and assistance provided by the Spanish Supercomputing Network (Red Espa\~nola de Supercomputaci\'on), as well as the computer resources used: the LaPalma Supercomputer, located at the Instituto de Astrof\'isica de Canarias.

G.\,H.\,acknowledges financial support from the Polish National Science Center (NCN), grant no. 2015/18/A/ST9/00578.  D.\,J. acknowledges support from the State Research Agency (AEI) of the Spanish Ministry of Science, Innovation and Universities (MCIU) and the European Regional Development Fund (FEDER) under grant AYA2017-83383-P.  DJ also acknowledges support under grant P/308614 financed by funds transferred from the Spanish Ministry of Science, Innovation and Universities, charged to the General State Budgets and with funds transferred from the General Budgets of the Autonomous Community of the Canary Islands by the Ministry of Economy, Industry, Trade and Knowledge.  This research was supported by the Erasmus+  programme of the European Union under grant number 2017-1-CZ01-KA203-035562.  L.\,N.~thanks the Natural Sciences and Engineering Research Council (Canada) for financial support through the Discovery Grants program.  Some computations were carried out on the supercomputers managed by Calcul Qu\'ebec and Compute Canada. The operation of these supercomputers is funded by the Canada Foundation for Innovation (CFI), NanoQu\'ebec, R\'eseau de M\'edecine G\'en\'etique Appliqu\'ee, and the Fonds de recherche du Qu\'ebec -- Nature et technologies (FRQNT).  J.\,A.~ thanks NSERC (Canada) for an Undergraduate Student Research Award (USRA).  D.\,J.\,S. acknowledges funding support from the Eberly Research Fellowship from The Pennsylvania State University Eberly College of Science. The Center for Exoplanets and Habitable Worlds is supported by the Pennsylvania State University, the Eberly College of Science, and the Pennsylvania Space Grant Consortium.  M.\,S.~acknowledges the financial support of the Operational Program Research, Development and Education -- Project Postdoc@MUNI (No. CZ.02.2.69/0.0/0.0/16\_027/0008360).

This project utilized data from the Digital Access to a Sky Century@Harvard (`DASCH') project at Harvard that is partially support from NSF grants AST-0407380, AST-0909073, and AST-1313370. This paper also makes use of the WASP data set as provided by the WASP consortium and services at the NASA Exoplanet Archive, which is operated by the California Institute of Technology, under contract with the National Aeronautics and Space Administration under the Exoplanet Exploration Program (DOI 10.26133/NEA9).

This project also makes use of data from the Kilodegree Extremely Little Telescope (`KELT') survey, including support from The Ohio State University, Vanderbilt University, and Lehigh University, along with the KELT follow-up collaboration.

\vspace{0.5cm}
\noindent
{\em Data availability}
\vspace{0.2cm}

\noindent
The {\em TESS} data used in this paper are available on MAST.  All other data used are reported in tables within the paper.  The {\tt MESA} binary evolution `inlists' are available on the {\em MESA} Marketplace: \url{http://cococubed.asu.edu/mesa_market/inlists.html}.

\bibliography{63328020_ver5.bib}

\end{document}